\documentclass[longauth]{aa}  
\usepackage{graphicx}
\usepackage{txfonts}
\usepackage{lipsum}
\usepackage[dvipsnames]{xcolor}
\usepackage{stfloats}
\usepackage{multirow}
\usepackage[version=4]{mhchem}
\usepackage[utf8]{inputenc}
\usepackage{booktabs}
\usepackage{amssymb}
\usepackage{caption} 
\usepackage{xhfill}
\usepackage{array}
\usepackage[euler]{textgreek}
\usepackage{orcidlink}

\newcolumntype{L}[1]{>{\raggedright\let\newline\\\arraybackslash\hspace{0pt}}m{#1}}
\newcolumntype{C}[1]{>{\centering\let\newline\\\arraybackslash\hspace{0pt}}m{#1}}
\newcolumntype{R}[1]{>{\raggedleft\let\newline\\\arraybackslash\hspace{0pt}}m{#1}}

\newcommand*{\dprime}{^{\prime\prime}\mkern-1.2mu}
\newcommand{\micron}{\textmu m\xspace}
\newcommand{\ditto}[1][.0pt]{\xrfill[.7ex]{#1}~\textquotedbl~\xrfill[.7ex]{#1}}

\usepackage{soul}

\begin{document} 

    \title{A JWST inventory of protoplanetary disk ices:}
    \subtitle{The edge-on protoplanetary disk HH~48~NE, seen with the Ice Age ERS program}
    \titlerunning{A JWST inventory of protoplanetary disk ices}

   \author{ J.~A.~Sturm              \inst{1}\thanks{sturm@strw.leidenuniv.nl}\orcidlink{0000-0002-0377-1316}       \and
            M.~K.~McClure            \inst{1}\orcidlink{0000-0003-1878-327X}   \and
            T.~L.~Beck               \inst{2}\orcidlink{0000-0002-6881-0574}   \and
            D.~Harsono               \inst{3}\orcidlink{0000-0001-6307-4195}   \and
            J.~B. Bergner            \inst{4}\orcidlink{0000-0002-8716-0482}   \and
            E.~Dartois               \inst{5}\orcidlink{0000-0003-1197-7143}   \and
            A.~C.~A.~Boogert         \inst{6}\orcidlink{0000-0001-9344-0096}   \and  
            J.~E.~Chiar              \inst{7,8}\orcidlink{0000-0003-2029-1549}   \and
            M.~A.~Cordiner           \inst{9,10}\orcidlink{0000-0001-8233-2436}   \and    
            M.~N.~Drozdovskaya       \inst{11}\orcidlink{0000-0001-7479-4948}   \and
            S.~Ioppolo               \inst{12}\orcidlink{0000-0002-2271-1781}   \and
            C.~J.~Law                \inst{13}\orcidlink{0000-0003-1413-1776}   \and
            H.~Linnartz              \inst{14}\orcidlink{0000-0002-8322-3538}   \and
            D.~C.~Lis                \inst{15}\orcidlink{0000-0002-0500-4700}   \and
            G.~J.~Melnick            \inst{13}                                  \and
            B.~A.~McGuire            \inst{16,17}\orcidlink{0000-0003-1254-4817}   \and
            J.~A.~Noble              \inst{18}\orcidlink{0000-0003-4985-8254}   \and
            K.~I.~\"Oberg            \inst{13}\orcidlink{0000-0001-8798-1347}   \and   
            M.~E.~Palumbo            \inst{19}\orcidlink{0000-0002-9122-491X}   \and
            Y.~J.~Pendleton          \inst{20}\orcidlink{0000-0001-8102-2903}   \and
            G.~Perotti               \inst{21}\orcidlink{0000-0002-8545-6175}   \and
            K.~M.~Pontoppidan        \inst{2}\orcidlink{0000-0001-7552-1562}   \and
            D.~Qasim                 \inst{22}\orcidlink{0000-0002-3276-4780}   \and
            W.~R.~M.~Rocha           \inst{1,14}\orcidlink{0000-0001-6144-4113}   \and
            H.~Terada                \inst{23,24}\orcidlink{0000-0002-7914-6779}   \and
            R.~G.~Urso               \inst{19}\orcidlink{0000-0001-6926-1434}   \and
            E.~F.~van~Dishoeck       \inst{1,25}\orcidlink{0000-0001-7591-1907}  
          }

    \institute{
    Leiden Observatory, Leiden University, P.O. Box 9513, NL-2300 RA Leiden, The Netherlands                                                \and
    Space Telescope Science Institute, 3700 San Martin Drive, Baltimore, MD 21218, USA                                                      \and
    Institute of Astronomy, Department of Physics, National Tsing Hua University, Hsinchu, Taiwan                                           \and
    Department of Chemistry, University of California, Berkeley, California 94720-1460, United States                                       \and
    Institut des Sciences Mol\'eculaires d’Orsay, CNRS, Univ. Paris-Saclay, 91405 Orsay, France                                             \and
    Institute for Astronomy, University of Hawai'i at Manoa, 2680 Woodlawn Drive, Honolulu, HI 96822, USA                                   \and
    Physical Science Department, Diablo Valley College, 321 Golf Club Road, Pleasant Hill, CA 94523, USA                                    \and
    SETI Institute, 339 N. Bernardo Ave Suite 200, Mountain View, CA 94043                                                                  \and
    Astrochemistry Laboratory, NASA Goddard Space Flight Center, 8800 Greenbelt Road, Greenbelt, MD 20771, USA                              \and
    Department of Physics, Catholic University of America, Washington, DC 20064, USA
                  \and
    Center for Space and Habitability, Universit\"at Bern, Gesellschaftsstrasse 6, 3012, Bern, Switzerland                                  \and
    Center for Interstellar Catalysis, Department of Physics and Astronomy, Aarhus University, Ny Munkegade 120, Aarhus C 8000, Denmark     \and
    Center for Astrophysics \textbar\ Harvard \& Smithsonian, 60 Garden St., Cambridge, MA 02138, USA                                       \and
    Laboratory for Astrophysics, Leiden Observatory, Leiden University, PO Box 9513, 2300 RA Leiden, The Netherlands                        \and
    Jet Propulsion Laboratory, California Institute of Technology, 4800 Oak Grove Drive, Pasadena, CA, 91109, USA                           \and
    Department of Chemistry, Massachusetts Institute of Technology, Cambridge, MA 02139, USA                                                \and
    National Radio Astronomy Observatory, Charlottesville, VA 22903, USA                                                                    \and
    Physique des Interactions Ioniques et Mol\'{e}culaires, CNRS, Aix Marseille Univ., 13397 Marseille, France                              \and
    INAF – Osservatorio Astrofisico di Catania, via Santa Sofia 78, 95123 Catania, Italy                                                    \and
    Department of Physics, University of Central Florida, Orlando, FL 32816, USA                                                            \and
    Max Planck Institute for Astronomy, K{\"o}nigstuhl 17, D-69117 Heidelberg, Germany                                                      \and
    Southwest Research Institute, San Antonio, TX 78238, USA                                                                                \and
    TMT International Observatory, 100 W Walnut St, Suite 300, Pasadena, CA USA                                                             \and
    National Astronomical Observatory of Japan, National Institutes of Natural Sciences (NINS), 2-21-1 Osawa, Mitaka, Tokyo 181-8588, Japan \and 
    Max-Planck-Institut f{\"u}r extraterrestrische Physik, Giessenbachstra{\ss}e 1, 85748 Garching bei M\"unchen, Germany 
    }

    \date{Received XXX; accepted YYY}
    \abstract{
    Ices are the main carriers of volatiles in protoplanetary disks and are crucial to our understanding of the protoplanetary disk chemistry that ultimately sets the organic composition of planets. 
    The Director’s Discretionary-Early Release Science (DD-ERS) program Ice Age on the James Webb Space Telescope (JWST) follows the ice evolution through all stages of star and planet formation. 
    JWST's exquisite sensitivity and angular resolution uniquely enable detailed and spatially resolved inventories of ices in protoplanetary disks.
    JWST/NIRSpec observations of the edge-on Class II protoplanetary disk HH~48~NE reveal spatially resolved absorption features of the major ice components H$_2$O, CO$_2$, CO, and multiple weaker signatures from less abundant ices NH$_3$, OCN$^-$, and OCS.
    Isotopologue $^{13}$CO$_2$ ice has been detected for the first time in a protoplanetary disk.
    Since multiple complex light paths contribute to the observed flux, the ice absorption features are filled in by ice-free scattered light. 
    This implies that observed optical depths should be interpreted as lower limits to the total ice column in the disk and that abundance ratios cannot be determined directly from the spectrum.
    The $^{12}$CO$_2$/$^{13}$CO$_2$ integrated absorption ratio of 14 implies that the $^{12}$CO$_2$ feature is saturated, without the flux approaching 0, indicative of a very high CO$_2$ column density on the line of sight, and a corresponding abundance with respect to hydrogen that is higher than ISM values by a factor of at least a few.
    Observations of rare isotopologues are crucial, as we show that the $^{13}$CO$_2$ observation allows us to determine the column density of CO$_2$ to be at least 1.6$~\times~$10$^{18}$~cm$^{-2}$; more than an order of magnitude higher than the lower limit directly inferred from the observed optical depth.
    Spatial variations in the depth of the strong ice features are smaller than a factor of two.
    Radial variations in ice abundance, e.g., snowlines, are significantly modified since all observed photons have passed through the full radial extent of the disk.
    CO ice is observed at perplexing heights in the disk, extending to the top of the CO-emitting gas layer. 
    Although poorly understood radiative transfer effects could contribute, we argue that the most likely interpretation is that we observe some CO ice at high temperatures, trapped in less volatile ices like H$_2$O and CO$_2$.    
    Future radiative transfer models will be required to constrain the physical origin of the ice absorption and the implications of these observations for our current understanding of disk physics and chemistry.
    }

    \maketitle
    
\section{Introduction}\label{sec:Introduction}
Life, as we know it, is largely formed from the elements C, H, O, N, and S (CHONS), which, in protoplanetary disks, are carried mostly as ices or organics on dust grains \citep{Henning2013chemistry,Furuya2013,Krijt2022_PPVII}. 
These grains are accreted by newborn planets, either as they form or during a later delivery by planetesimals residing in warm or cold debris reservoirs (asteroid and Kuiper Belt analogs) in more mature systems.
Such organic-rich icy grains are formed in molecular clouds and are funneled into the planet-forming regions of disks via an infalling protostellar envelope during the early stages of star formation \citep[e.g.,][]{Visser2009,Hincelin2013,Drozdovskaya2016}. 
During infall, some of the ices sublimate during episodic, protostellar mass accretion events \citep{Kim2012}, and they could dissociate to their atomic constituents in the gas phase before reforming and freezing out again in the disk, partly resetting the disk's ice chemistry relative to the cloud chemistry \citep{Ballering2021}. 
The relative amount of different ice species inherited by the disk from the cloud versus reset during this process is observationally still poorly constrained  \citep{ObergBergin2021}.

Within disks, the ice species are expected to be vertically and radially segregated by 3D `snowlines', which is the point where species sublimate from the ice phase. 
Grain growth and gravitational settling produce a vertical dust distribution that dictates where the stellar radiation is absorbed and scattered, setting a strong vertical temperature gradient. 
A corresponding chemical gradient arises between the disk midplane, where planetesimals form from large, icy grains, and the disk's upper layers, where UV radiation photodesorbs the ice from the grain surface at a vertical snowsurface \citep{Dominik2005,kamp2018}. 
Similarly, a radial temperature gradient gives rise to snowlines at certain distances from the central star, where inherited or reformed ices sublimate. 
Disk snowlines gave rise to Solar System planets, moons, and small bodies with different compositions \citep{bergin2015} and may have also set the bulk atmospheric properties of the planets \citep{Oberg2011,Madhusudhan2019}. 
The distribution of major CHONS ices during the initial stages of planet formation in protoplanetary disks can be inferred from comets, asteroids and icy moons in the Solar System (e.g., with Rosetta, Hayabusa 2, OSIRIS-REx, Juice) or in situ snapshots of ices at different stages of planet formation. 
The latter, in particular, is critical to understand the variety of processes and planetary outcomes during planet formation \citep{Krijt2022_PPVII}.

Individual ice species are detectable in the laboratory through fundamental vibrational modes seen in the infrared between 1 and 140~\micron \citep[e.g.,][and references therein]{Rocha2022}. 
These laboratory spectra can be used to identify astrophysical ices in different environments. 
Ices have been detected in absorption against the hot continuum provided by stars located behind cold, dense molecular clouds, in protostellar envelopes, and in the cold outer regions of edge-on protoplanetary disks \citep{Boogert2015}, and in emission from the midplanes of a handful of disks around T Tauri and Herbig~Ae/Be stars from 44~--~63~\micron with the Infrared Space Observatory (ISO) and the \textit{Herschel} Space Observatory \citep{Chiang2001,mcclure2012,mcclure2015,min2016}. 
Ice emission requires full radiative transfer modeling to fully interpret, but for pure absorption spectroscopy, it has been standard practice to calculate logarithmic ice optical depths from spectroscopic data using:
\begin{equation}
    \label{eq:optical depth}
    \tau = -\mathrm{ln}\left(\frac{F_{\nu}}{F_{\nu\mathrm{,cont}}}\right),
\end{equation}
where $F_{\nu}$ is the observed flux and $F_{\nu\mathrm{,cont}}$ is the flux of the fitted continuum. 
Then lab-measured band strengths of the ice features are used to convert the observed optical depths to column densities by: 
\begin{equation}\label{eq:column density from band strength}
   N = \frac{\int\tau_\nu \mathrm{d}\nu}{A},
\end{equation}
where $\nu$ is wavenumber in cm$^{-1}$ and $A$ is the band strength in cm~molecule$^{-1}$. 
If the ice features could be assumed to come from the same absorbing region, then relative ice abundances could be derived from column densities in a model-independent way. 
In the past, most of the available ice spectra consisted of spatially unresolved spectroscopy of point sources; therefore, the simplifying assumptions of pure ice absorptions within the same beam seemed appropriate for ices seen towards background stars and protostars. 
In this manner, ice abundances relative to \ce{H_2O} have been derived in a few molecular clouds using lines of sights towards background stars, many protostars, and a small handful of edge-on disks (EODs) using the Very Large Telescope (VLT) or the {\it Spitzer} Space Telescope. 

This assumption was known to be incorrect for ices in any source with an extended physical structure, like a disk. 
However, recent work has suggested that this traditional assumption of pure ice absorption may not hold in most astrophysical environments relevant to star formation.
When icy mantles grow to micrometer sizes, they are seen through a combination of both absorption and scattering, which modifies their feature profiles and may strongly impact how well column densities translate into relative abundances, even towards targets with no source structure like stars behind molecular clouds \citep{Dartois2022}.
This effect would be expected to be enhanced towards sources with internal structure, like protostars or disks. 

For disks in particular, the observations themselves are challenging. 
First, ices are concentrated in the midplane, which only becomes optically thin in the far-infrared.
In the absence of space-based far-infrared telescopes, the only current option to observe ices is to probe them in the mid-infrared, although this samples ices in the disk atmosphere rather than in the disk midplanes, which is optically thick in the infrared. 
At these wavelengths, ices have been detected by broadband, coronagraphic ground-based scattered light imaging of sufficiently bright, face-on disks, e.g., the Herbig~Ae stars HD~142527 and HD~100546 \citep{Honda2009,Honda2016}, and through spectroscopy of edge-on systems ($i$~$>$~70\degr).
In the latter case, the optically thick midplane acts as a coronagraph to block thermal emission from the central star and hot dust in the disk's midplane at 1-5~au, reducing the contrast enough to detect the signatures of cold ices beyond 20~au in absorption against the faint light from the star or inner dust rim scattered through the disk's atmosphere \citep{Pontoppidan2007, PaperII}. 
This technique allows the measurement of multiple ice species simultaneously, which is critical for determining CHONS abundances.

Previous spectroscopic studies of edge-on disks were limited by their technical capabilities and sample selection. 
Unresolved, single-slit ground-based spectroscopy detected water ice in a handful of disks, including CRBR~2422.8-3423, HV~Tau~C, HK~Tau~B, a couple of Orion silhouette disks, and the disks around the Herbig~Ae/Be stars PDS~144~N and PDS~453 \citep{Thi2002,Terada2007,Terada2012,Terada2017}. 
However, detection of the main carbon dioxide (\ce{CO2}) and carbon monoxide (CO) ice features are limited by absorption from Earth's atmosphere and the sky background at 4~--~5~\micron. 
Additional observations from space of edge-on disks ASR41, 2MASS~J1628137-243139, HV~Tau, HK~Tau, UY~Aur, and IRAS~04301 with AKARI \citep{Aikawa2012} and CRBR~2422.8-3423 with \textit{Spitzer}~IRS \citep{Pontoppidan2005} confirmed the ground-based water detections and added CO$_2$ detections for the latter five sources. 

However, the last two systems demonstrate the second difficulty with ice measurements in disks: target selection. 
Without ancillary data, it can be difficult to disentangle whether the observed ice bands originate in the disk itself or in other foreground material (e.g., in a surrounding cloud or a protostellar envelope). 
Selecting samples of mature, envelope-free disks should solve this problem, but historically it has been difficult to identify edge-on disks a priori. 
Extinction, including self-extinction by their outer disks and scattering from the inner edge, can result in edge-on disks being mistaken for younger protostars due to their rising shape in the mid-infrared when identified purely based on spectral energy distributions (SEDs) \citep[][Figure 3]{Robitaille2006, McClure2010}. 
Only recently, small samples of non-embedded edge-on disks have been compiled and resolved at multiple wavelengths, using optical scattered light and resolved millimeter imaging \citep{Villenave2020}. 
For these validated disks, the sensitivity, wavelength coverage, spectral resolution, and spatial resolution of the \textit{James Webb} Space Telescope's (JWST) near and mid-infrared spectrographs, NIRSpec and MIRI, now allow the 2D spatial distribution of ices in such disks to be mapped with a 0\farcs{1} to 0\farcs{3} pixel scale. 

As part of the JWST Early Release Science Program, Ice Age, we observed the nearly edge-on \citep[$i$~=~$\sim$83\degr][]{PaperI} protoplanetary disk HH~48~NE around a K7 star in the Chamaeleon~I star-forming region \citep[185~pc,][]{GAIADR3} that has an estimated foreground-cloud visual extinction of only $A_\mathrm{V}$~=~5.
A flared disk of 1.3~--~1.7$\dprime$ diameter (120-160~au radius) was seen around HH~48~NE by the \textit{Hubble} Space Telescope (HST) \citep{Stapelfeldt2014} and by the Atacama Large (sub-)Millimeter Array (ALMA) \citep{Villenave2020,PaperI}, a size that is well-suited to the JWST/NIRSpec instrument's field of view. 
In the current study, we present the first results of these JWST/NIRSpec observations of the ices in the outer regions of HH~48~NE.

This paper describes the first complete gas and ice inventory of an edge-on protoplanetary disk observed with JWST/NIRSpec in a model-independent way. 
Due to the richness and novelty of these data, new science can be gleaned directly from the data prior to radiative transfer fitting. 
We discuss the interpretation of disk-integrated and spatially resolved ice features, finding that multiple scattering from many regions in the disk makes the traditional metrics of optical depths and column densities meaningless. 
This highlights the ultimate necessity of radiative transfer for determining ice abundances in protoplanetary disks. 

The paper is organized as follows: We first present the observations in Sect.~\ref{sec:Observations}. 
Then, in Sect.~\ref{sec:Results}, we provide details on the detected gas and ice features in the disk-integrated spectrum, along with the spatial variations of the ice feature depth. 
In Sect.~\ref{sec:Discussion}, we discuss the implications of the observations and pose unresolved questions for future research. 
Finally, Sect.~\ref{sec:Conclusion} summarizes our findings and provides concluding remarks.

\section{Observations}\label{sec:Observations}
The protoplanetary disk around HH~48~NE was observed with NIRSpec onboard JWST on 2022 July 19th, as part of the Director’s Discretionary-Early Release Science (DD-ERS) program ``Ice Age: Chemical evolution of ices during star formation'' (ID 1309, PI: McClure). 
The spectroscopic data were taken with a standard 4-point dither pattern for a total integration of 2451~s, with the pointing centered on 11$^\mathrm{h}$04$^\mathrm{m}$23.56$^\mathrm{s}$, -77\degr18$^\prime$07.26$\dprime$. 
The disperser-filter combination used was G395H-F290LP, which covers wavelengths from 2.87~--~4.08~\micron and 4.19~--~5.25~\micron at a resolving power of $R$ $\sim$2700. 
Because the target is extended, target acquisition was not carried out, and HH~48~NE was blind-pointed into the NIRSpec IFU aperture for spatially resolved imaging spectroscopy across the disk and inner outflow extent.  
A second pointing was observed at an offset of 3 arcminutes south-east of HH~48~NE to subtract the astrophysical background from the target flux.  

The data were processed through the November 2022 development version of the JWST NIRSpec pipeline (2022\_2a; 1.8.3.dev26+g24aa9b1d), which included a first fix for a pronounced `ringing' throughput structure seen in the previous version of the calibrated three dimensional (3D) IFU cube product for NIRSpec.  
Calibration reference file database version 11.16.16 was used, which included the updated onboard flat field and throughput calibrations for absolute flux calibration accuracy estimate on the order of $\sim$10\%.  
The standard steps in the JWST pipeline were carried out to process the data from the 3D ramp format to the cosmic ray corrected slope image.  
Further processing of the 2D slope image for WCS, flat fielding, and flux calibration were also done using standard steps in the ``Level 2'' data pipeline calwebb\_spec2.  
To build the calibrated 2D IFU slice images into the 3D datacube, the ``Stage 3'' pipeline was run step-wise, and intermediate products were investigated for accuracy.  
The four dithers were combined using the ``drizzle'' algorithm with equal weighting and full pixel regions used.  
The final pipeline processed product presented here was built into 3D with the outlier bad pixel rejection step turned off, as running this over-corrected and removed target flux. 
As a result, clusters of bad pixels are seen in some velocity channels of the IFU cube that are flagged by hand when necessary based on the 2D spatial information. 

\section{Results}\label{sec:Results}
Figure~\ref{fig:rgbim} presents the spatially-resolved continuum image of HH~48~NE at 4~\micron together with the extended \ce{H2} 0-0 S(9) emission at 4.7~\micron and the jet traced by [O~I] at 0.6~\micron with HST \citep{Stapelfeldt2014}.
The cube is rotated by -10\degr\ using \texttt{scipy} cubic spline interpolation to align the major axis of the disk horizontally for visualization purposes.
The total field of view (FOV) of the NIRSpec instrument is $3\dprime~\times~3\dprime$, which means that HH~48~SW \citep[located at 2\farcs{3};][]{PaperI} lies outside of the field, but some hints of gas emission from HH~48~SW are visible towards the southwest (see Fig.~\ref{fig:rgbim}).
The HH~48~NE protoplanetary disk appears spatially resolved in the NIRSpec spectral cube, with a major axis of 1\farcs{6} (296~au) and a minor axis of 1\farcs{1} (204~au) at 4~\micron determined using a 5~$\sigma$ clip.
The spatial resolution of the observations varies from 0\farcs{09} at 3~\micron to 0\farcs{15} at 5~\micron (17--28~au), which implies that the spaxels ($0.1~\dprime~\times~0.1\dprime$) are slightly smaller than the size of the effective PSF at the longest wavelengths.
The continuum is asymmetric in the lower surface, extending further towards the west than to the east (cf. the same continuum on a log scale in the top left panel of Fig.~\ref{fig:spatial dependence main ices}). 
This pattern is consistent with the optical HST observations \citep{Stapelfeldt2014,PaperI} and may be a result of gravitational interaction between the two sources.

A 1D, disk-integrated spectrum is extracted within a rectangle of 1\farcs{6}~$\times$~1\farcs{1} around the disk, centered on 11$^\mathrm{h}$04$^\mathrm{m}$23.38$^\mathrm{s}$, -77$^\mathrm{o}$18$^\prime$06.46$\dprime$ (see Fig.~\ref{fig:rgbim}) by summing up the flux over the spaxels with $S/N$>5.
We estimate the noise level of the disk-integrated spectrum in a flat spectral region without gas and ice features (3.85~--~4~\micron), which results in a local standard deviation of $\sigma$~=~0.027~mJy, or a $S/N$ of $\sim$110 on the continuum.
We present the disk-integrated NIRSpec spectrum of HH~48~NE in the top panel of Fig.~\ref{fig:overview_figure}. 
The spectrum shows a fairly flat continuum of $\sim$3~mJy, which is consistent with the broadband Spitzer/IRAC fluxes calculated using a Gaussian decomposition of the two binary components (HH~48~NE and SW, separated 2\farcs{3} from each other) by \citet{Dunham2016}.
This spectrum provides the total throughput of the stellar energy through the disk and has the highest achievable $S/N$ ratio on a spectrum in this data set for ice analysis.
Several gas emission lines and ice absorption features are detected, which we present in the subsequent sections.

\subsection{Gas inventory}
\begin{figure}[!t]
    \centering
    \includegraphics[width = \linewidth]{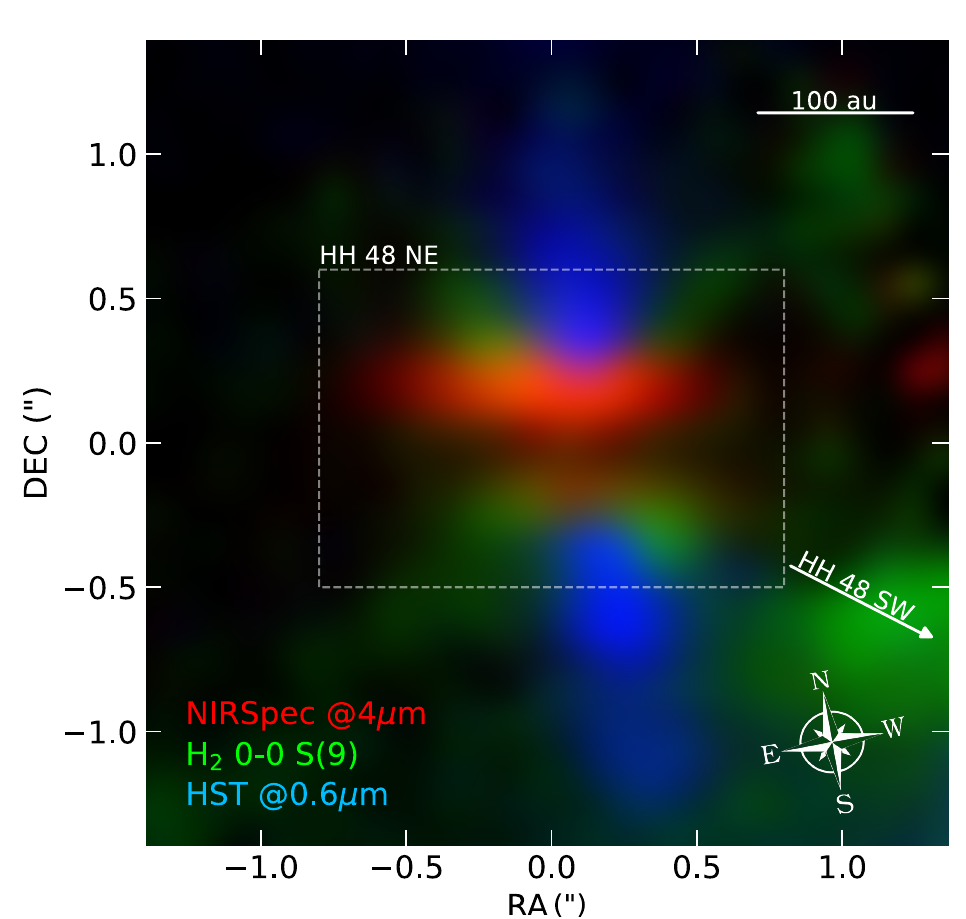}
    \caption{Overview of the different spatial components of the HH~48~NE system using a composition of observations. The JWST/NIRSpec disk continuum at 4~\micron is shown in red, the jet is shown in blue using HST/ACS observations of the [O~I] line at 0.6~\micron \citep{Stapelfeldt2014}, and the vertically extended disk wind is shown in green using the \ce{H2 0-0 S(9)} transition. The region used to extract a disk-integrated spectrum is shown by the dashed rectangle. The continuum in this region is shown on a logarithmic scale in Fig.~\ref{fig:spatial dependence main ices}. The direction towards HH~48~SW is marked with a white arrow.}
    \label{fig:rgbim}
\end{figure}

\begin{figure*}
    \centering
    \includegraphics[width = \textwidth]{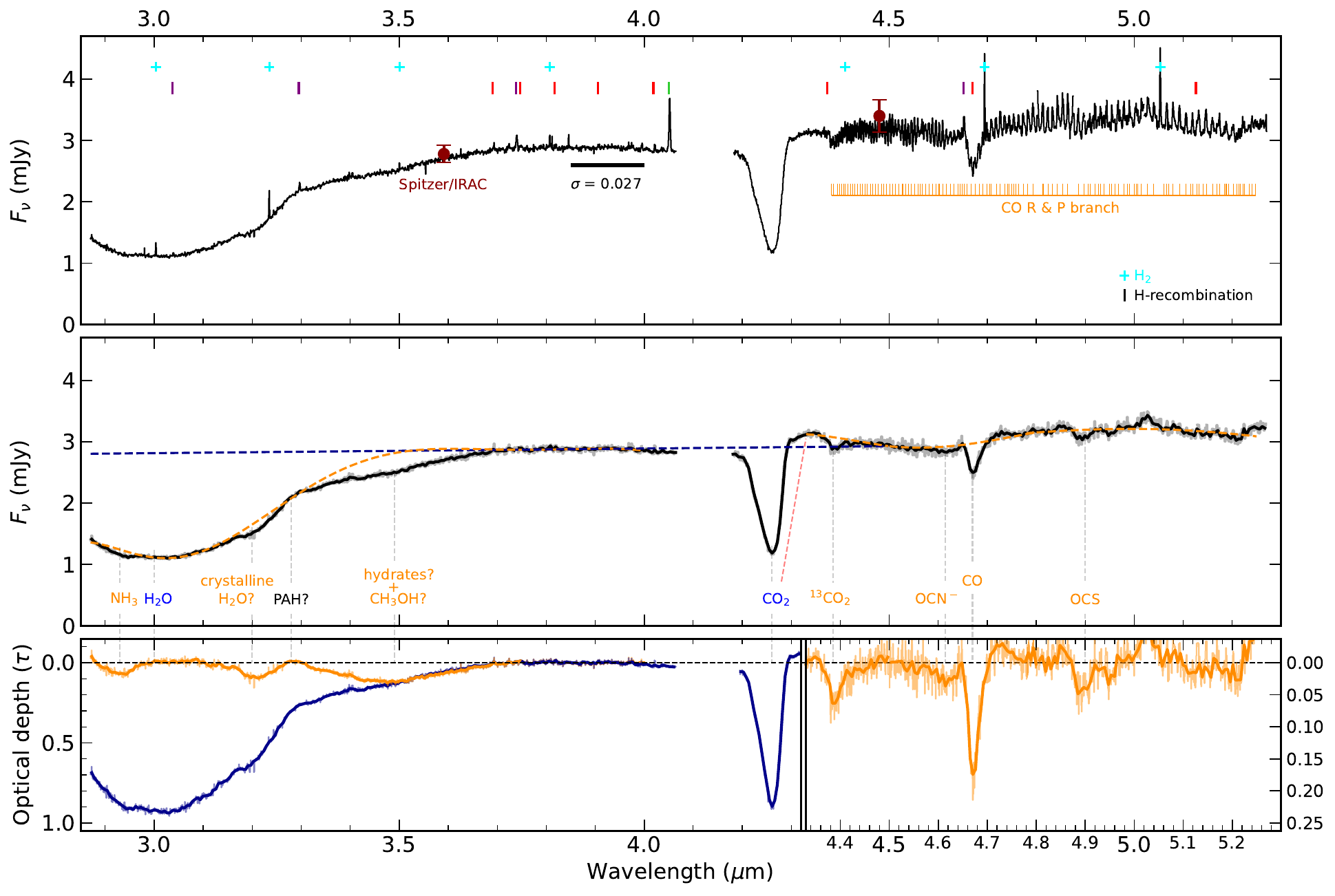}
    \captionsetup{format=hang}
    \caption{Overview of the disk-integrated JWST/NIRSpec spectrum of HH~48~NE. \\
    \textbf{Top panel:} Disk-integrated JWST/NIRSpec spectrum with the main gas lines labeled. The molecular hydrogen O and S series lines are marked in cyan. Bracket, Humphreys, and Pfund recombination series of atomic hydrogen are marked in green, red, and purple, respectively. The CO $R$ (4.3~--~4.7~\micron) and $P$ (4.7~--~5.2~\micron) branches of ro-vibrational transitions are marked in orange. The two literature Spitzer/IRAC data points are shown in dark red for comparison. The wavelength range 4.08~--~4.19~\micron is not covered in the used spectral setup.\\
    \textbf{Middle panel:} Disk-integrated JWST/NIRSpec spectrum with the gas lines removed in black.  The procedure for removing the CO lines is described in detail in Appendix~\ref{app:co fitting}. The positions of the main vibrational modes of the more abundant ice species are labeled, with the continua used to calculate their optical depths shown as dashed lines with corresponding colors. \\
    \textbf{Bottom panel:} Optical depth of the ice features in the disk-integrated spectrum shown in the middle panel. The two different colors represent optical depths for two different continuum fits, one for the \ce{H2O} and \ce{CO2} features in blue, and one for the weaker ice features neglecting the \ce{CO2} and \ce{H2O} features in orange. Both corresponding continua are shown as a dashed line in the middle panel.}
    \label{fig:overview_figure}
\end{figure*}

Ro-vibrational molecular emission lines and hydrogen recombination lines are ubiquitously found in protoplanetary disks, arising from the warm inner disk regions \citep{Pontoppidan2014}.
In edge-on disks, these - usually spatially unresolved - lines can be observed through photon scattering in a similar fashion as the thermal dust radiation.
Detected gas emission lines in HH~48~NE are labeled in the upper panel of Fig.~\ref{fig:overview_figure}.
Atomic hydrogen recombination lines are detected from the Bracket ($n$~=~4), Pfund ($n$~=~5), and Humphreys ($n$~=~6) series. 
These lines are spectrally unresolved and have a constant ratio with respect to the spatially resolved scattered continuum.
This indicates that the lines have a similar region of origin as the continuum emission in the inner disk, and the spatial extent of the emission is a result of photon-scattering in the outer disk.
In this spectral setting, there are 7 molecular hydrogen lines from the $v$~=~1--0 (O; $\Delta J$~=~-2) and $v$~=~0--0 (S;$\Delta J$~=~+2) series.
These lines have a slightly larger spatial extent than the disk continuum (see Fig.~\ref{fig:rgbim}), and likely result from a vertically extended disk wind or outflow cavity.


Additionally, we detect many ro-vibrational lines from the CO $R$ ($\Delta J$~=~1) and $P$ ($\Delta J$~=~-1) branches, up to $J$~=~48 (E$_{\rm up}$~=~10,056~K) and from several vibrationally excited states.
These lines are broader than expected based on the instrumental resolution, which means that they may trace a warm, high-velocity component in the inner disk ($v \sim$150~km~s$^{-1}$) or the wind-launching region.
The estimated line width is consistent with high resolution VLT/CRIRES observations of similar systems \citep[see e.g.,][]{Herczeg2011,Brown2013,Banzatti2015} that have a range of full width at half maximum (FWHM) of 50 -- 200~km~s$^{-1}$ which corresponds to an emitting radius of $\sim$0.05~au.
The spatial extent of the CO lines is similar to the continuum in the upper surface, but the contribution from the lower surface is missing (see Fig.~\ref{fig: CO spat appendix}). 
The origin of this phenomenon is unknown, but it could result from a difference in the vertical height of the CO-emitting layer compared to the continuum which could block part of the emission.

We removed the CO lines from the spectrum to accommodate a better fit of the continuum and identification of the ice absorption features, which are the main focus of this paper. 
For this purpose, we fit a simple LTE slab model to the lines with a constant temperature (2070~K) and column density (1.4$~\times~$10$^{-19}$~cm$^{-2}$), and subtracted it from the spectrum (see Appendix~\ref{app:co fitting}).
Since the spatial extent of the CO emission is similar to the continuum in the upper surface, with a constant flux ratio of 10 (see Fig.~\ref{fig: CO spat appendix}), we assume that the propagation of photons from CO molecules is similar to the continuum emission that originates in the central regions of the disk.
This implies that we detect the same CO gas component throughout the disk, and its spatial extent is the product of scattering to the outer disk.
Therefore, we scale the best-fit CO spectrum with a constant scaling factor and subtract it from every spaxel that contains dust emission.
Additionally, we flag the hydrogen lines and present the gas line subtracted spectrum in the middle panel of Fig.~\ref{fig:overview_figure}.

\subsection{Ice inventory}\label{ssec: Observations - ice features}
The middle panel of Fig.~\ref{fig:overview_figure} clearly reveals broad absorption features due to different ice species, after subtraction of the CO and other gas phase lines.
We fit the continuum around the main ice absorption features with a polynomial, using the wavelength ranges and fit orders specified in Table~\ref{tab:continuum points} and show the result of the continuum fit in the middle panel of Fig.~\ref{fig:overview_figure}. 
Since the spectrum does not include the full blue wing of the water ice feature, we fitted a separate linear continuum for the water (\ce{H2O}) feature that is extrapolated down to 2.87~\micron.
The ice spectrum is then converted to an ice optical depth\footnote{In edge-on disk observations, this should be interpreted as ``the logarithmic fraction of light at each wavelength that reaches cold regions in the disk with ice and is absorbed there'' instead of the usual ``logarithmic fraction of absorbed light at specific wavelengths'' with respect to a continuum background source. This implies that there is no straight-forward relationship with the column density.} using Eq. \ref{eq:optical depth}.
The optical depth indicates the fraction of light at each wavelength that is absorbed by the icy dust grains, on a logarithmic scale, which is linearly correlated with the ice column density encountered along the light path neglecting radiative transfer effects.
The resulting spectrum is shown in blue in the bottom panel of Fig.~\ref{fig:overview_figure}.
The integrated optical depths for the detected features are given in Table~\ref{tab:continuum points}, with the same band strengths as in \citet{McClure2023} for consistency.
These values cannot be used to calculate column densities directly due to the flux contributions of multiple light paths through the disk, as we will show in Sect. \ref{ssec:Discussion - column densities}.
We would like to note that the exact choice for the continuum baseline can result in an uncertainty of up to 10\% in integrated optical depth.

We describe the major observed ice features in the next subsections.
The shape and depth of ice features can be significantly altered by radiative transfer and grain growth effects, especially in scattering-dominated edge-on disks \citep[see e.g.,][]{Dartois2022}, which complicates the analysis of the ice bands. 
A detailed study of the ice composition, environment, and abundance requires radiative transfer modeling is therefore left for a forthcoming paper.

\begin{table*}
    \caption{Properties of the detected ice features in the disk-integrated spectrum of HH~48~NE.}
    \label{tab:continuum points}
    \begin{tabular}{L{.1\linewidth}L{0.1\linewidth}L{.34\linewidth}L{0.08\linewidth}L{0.09\linewidth}L{0.15\linewidth}}
        \bottomrule
        \toprule
         Species       &$\lambda$ (\micron)   & Continuum wavelength range (\micron)           & Fit order & $\int \tau_\nu$d$\nu$ & $A$  \\
         &&&&(cm$^{-1}$)&(cm molecule$^{-1}$)\\
         \midrule
         \ce{H2O}          &3.0  & 3.7 - 4, 4.42 - 4.58                               & 1    & 381$^{1.}$   &  2.0 $~\times~$  10$^{-16}$ \hfill $^{(1)}$\\
         \ce{CO2}          &4.26 & 3.7 - 4, 4.42 - 4.58, 4.71~-~4.87, 4.95~-~5.25     & 5    & 23.0  &  1.1 $~\times~$ 10$^{-16}$ \hfill $^{(2)}$\\
         \ce{CO}           &4.67 & \ditto                                             & \textquotedbl  & 2.0   &  1.1 $~\times~$ 10$^{-17}$ \hfill $^{(1)}$\\         
         \ce{OCN^-}        &4.6  & \ditto                                             & \textquotedbl  & 0.7   &  1.3 $~\times~$ 10$^{-16}$ \hfill $^{(3)}$\\
         \ce{OCS}          &4.9  & \ditto                                             & \textquotedbl  & 0.6   &  1.2 $~\times~$ 10$^{-16}$ \hfill $^{(4)}$\\
         \ce{NH3}          &2.93 & 2.86~-~2.9, 2.98~-~3.15, 3.27~-~3.31, 3.7~-~4.0    & 7    & 4.3   &  1.7 $~\times~$ 10$^{-16}$ \hfill $^{(5)}$\\
         \ce{NH3\cdot H2O} &3.5  & \ditto                                             & \textquotedbl  & 27.3  &  2.6 $~\times~$ 10$^{-17}$ \hfill $^{(6)}$\\
         \ce{^13CO2}       &4.39 & 3.65~-~4, 4.32~-~4.36, 4.43~-~4.5                  & 5    & 1.0   &  7.1 $~\times~$ 10$^{-17}$ \hfill $^{(1)}$\\
         \midrule
    \end{tabular}
   \\Notes: 1. This is the integrated absorption starting from 2.87~\micron, 
   \\References: (1) \citet{Gerakines1995}, (2) \citet{Gerakines2015}, (3) \citet{vanBroekhuizen2005}, (4) \citet{Yarnall2022}, (5) \citet{Hudson2022}, (6) \citet{Dartois2001}
\end{table*}

\subsubsection{\ce{H2O} and \ce{NH3}: 3~\micron}\label{sssec:ice inventory - water}
The 3.0~\micron absorption feature is primarily attributed to the \ce{H2O} OH stretching mode and is similar to previous spectra observed towards background stars with JWST.
The feature exhibits a well-defined peak with an optical depth of 0.9 and a relatively smooth shape, displaying a peak at 3.0~\micron and an absorption wing on the red side. 
The NIRSpec filters used do not capture the blue wing of the feature below 2.8~\micron. 
Analyzing the red wing is complicated by a potential weak emission feature of polycyclic aromatic hydrocarbons (PAH) at 3.28~\micron \citep{Draine1984}. 
A comparison with PAH features at MIRI wavelengths, that are strongly correlated with this feature \citep{Draine1984}, is required to determine its actual contribution to the continuum. 
Currently, a continuum is fitted avoiding potential weak absorption features within the OH stretch mode of water using the wavelength ranges specified in Table~\ref{tab:continuum points}, illustrated as the orange line in the middle panel of Fig.~\ref{fig:overview_figure}.

Other features of the 3~\micron feature can be ascribed to a combination of additional species and radiative transfer effects\footnote{In the case of a protoplanetary disk the effect of radiative transfer on the feature shape includes contributions from the phase-dependent scattering opacity of the grains \citep[see e.g.,][]{Dartois2022}, but also includes the fact that only specific paths through the disk are viable and the fact that the absorption feature may be saturated locally in the disk.}.
The narrow feature centered at 2.93~\micron is attributed to the ammonia NH stretch mode \citep{Dartois2002}. 
Its position and shape are uncertain since it is located right on the blue edge of the filter, which further complicates the exact location of the continuum.
The feature at 3.2~\micron is similar to those reported for the silhouetted disk 216-0939 in the M43 region and the edge-on disk around the Herbig star PDS 453 \citep{Terada2012a,Terada2017}, and is interpreted as a result of large-particle-size ($\sim$1~\micron) crystallized water ice absorption.
It is known that both larger grains and a higher degree of crystallization can cause a shift from the typical wavelength of this feature (3.1~\micron) to 3.2~\micron \citep{Smith1989}.
The overall strength compared to the peak at 3~\micron is a factor of 2~--~3 lower than in the former sources, which suggests that the fraction of crystallized \ce{H2O} ice is lower in HH~48~NE.
The broad feature at $\sim$3.5~\micron is observed in multiple background/protostellar sources, correlating with the water ice absorption \citep[e.g.][]{Brooke1996, Brooke1999, Dartois2002, Shimonishi2016}.
This absorption feature is often attributed to ammonia hydrates, implying an OH stretching mode absorption shifted by interaction with the nitrogen atom of ammonia embedded in water ice. 
The appearance and optical depth of this mode is consistent with the detection of the NH stretch at around 2.97~\micron for ammonia. 
Some potential additional contributions by CH stretching modes have also been presumed in this range, based on laboratory experiments using H atom irradiated carbon grains \citep{Mennella_2010}. 
Although the distinct, narrow peak of \ce{CH3OH} at 3.54~\micron is absent \citep[see Fig.~9 in][]{McClure2023}, a small contribution of \ce{CH3OH} cannot be excluded based solely on this feature and will be constrained better with the combined MIRI and NIRSpec data. 
Unfortunately, the less abundant water isotopologue, HDO, with a feature at 4.07~\micron is not within the observed wavelength range, due to the NIRSpec G395H detector gap, and is therefore excluded from the analysis.

\subsubsection{\ce{CO2} and \ce{^13CO2}: 4.2~--~4.4~\micron}\label{sssec:ice enventory - co2}
The carbon dioxide (\ce{CO2}) absorption feature at 4.26~\micron is detected with a similar peak optical depth as the \ce{H2O} feature.
The peak position of the observed \ce{CO2} stretching mode band is shifted compared to pure \ce{CO2} (4.27~\micron). 
This shift can be attributed to \ce{CO2} in  polar ice environments or by ices exposed to energetic processing at low temperatures \citep{Ioppolo2013,Ioppolo2022}. 
The main isotopologue has a red wing in emission (see red dashed line in the middle panel of Fig.~\ref{fig:overview_figure}), which is likely caused by a phase scattering effect at a high degree of disk inclination involving the scattering opacity of large (>1~\micron) grains \citep{Dartois2022}. 
The \ce{^13CO2} isotopologue absorption band is located in this scattering wing, so for that reason we fitted the continuum for this feature separately with a local fit specified in Table~\ref{tab:continuum points}. 
A local fit introduces further uncertainty in the derived column density, but it is necessary here as the main \ce{CO2} feature contributes to the continuum of the \ce{^13CO2} feature.

As for \ce{^12CO2}, the peak of the \ce{^13CO2} feature is shifted relative to the lab-measured peak of 4.38~\micron of pure \ce{^13CO2} to 4.39~\micron. 
\citet{Boogert2000} showed that this peak shift can be ascribed to absorption by \ce{^13CO2} ice mixed with molecules with a large dipole moment, for example, \ce{H2O} and \ce{CH3OH}, at high concentrations.
The derived \ce{^12CO2}/\ce{^13CO2} ratio is low, with a ratio in peak optical depth of 14 and a ratio in the integrated optical depth of 22.
Assuming that both features arise from the same region in the disk, and using the band strengths reported in \citet{Gerakines1995} and \citet{Gerakines2015} for \ce{^12CO2} and \ce{^13CO2}, respectively, we find a molecular ratio of 14 compared to the local interstellar medium (ISM) ratio of 77 \citep{Wilson1999_13Cratio} and the local dark cloud ratio of 69~--~87 \citep{McClure2023}.
This low \ce{^12CO2}/\ce{^13CO2} ratio indicates that the main isotopologue may be saturated\footnote{Saturation in the context of scattered light implies that the strength of the absorption (or optical depth) is dominated by the fraction of light that makes it through icy, cold regions of the disk, instead of the column density of ice along the light path. In protoplanetary disks, saturation happens without the observed flux approaching 0.} at certain positions along the line of sight, a point to which we return in Sect. \ref{ssec:Discussion - column densities}.

\subsubsection{\ce{CO}, \ce{OCN^-}, and \ce{OCS}: 4.4~--~5~\micron}\label{sssec:ice enventory - co}
The right hand side of the bottom panel of Fig.~\ref{fig:overview_figure} shows the spectrum in the range 4.32~--~5.24~\micron, where features attributed to CO at 4.67~\micron and \ce{^13CO2} at 4.39~\micron are detected. 
The main CO stretch feature at 4.67~\micron is detected at a peak optical depth of 0.2. 
The feature shows signs of a potential red wing in emission, very similar to that observed for the \ce{CO2} feature. 
The absorption feature appears asymmetric and the total observed FWHM of the feature (12.4~cm$^{-1}$) is also broader than expected from pure CO \citep[3~cm$^{-1}$;][]{Palumbo2006}.
This may indicate the presence of an additional component of a CO mixture with \ce{CO2} or \ce{H2O}, similar to \citet{Pontoppidan2003} and \citet{Oberg2011_Spitzer_legacy}.
Meanwhile, \ce{^13CO} is not detected, but the 3$\sigma$ upper limit obtained ($\tau$ $<$0.03) results in a \ce{^12C}/\ce{^13C} ratio lower limit of 6, which is consistent with the observed ratio of 14 for \ce{CO2}.

The observed CO ro-vibrational gas lines produce in some places a quasi-continuum by line overlap.
We have corrected for this effect as much as possible by subtracting a slab model (see Appendix~\ref{app:co fitting}), but the remaining systematic error of this method is hard to determine.
Residual weak ice absorption should therefore be treated with care. 
The feature at 4.61~\micron is consistent with lab measurements of the cyanate anion \ce{OCN^-} \citep[e.g.,][]{Grimgreenberg1987, Hudson2001}. 
This feature is observed in a variety of sources, including low- and high-mass YSOs \citep[e.g., ][]{pendleton1999, vanbroekhuizen2004, Boogert2015} and it is consistent with the \ce{OCN^-} band observed in laboratory analogues after UV-photolysis, ion bombardment, and thermal processing of various icy mixtures \citep[e.g., ][]{palumbo2000a, vanbroekhuizen2004}.
The feature centered at 4.9~\micron is consistent with OCS ice. 
A comparison with laboratory data suggests that the feature arises from OCS embedded in mixed ices and in regions of the disk at different temperatures \citep{Palumbo1995, Palumbo1997, Ferrante2008, Garozzo2010, Boogert2022}.

\begin{figure*}
    \centering
    \includegraphics[width = \linewidth]{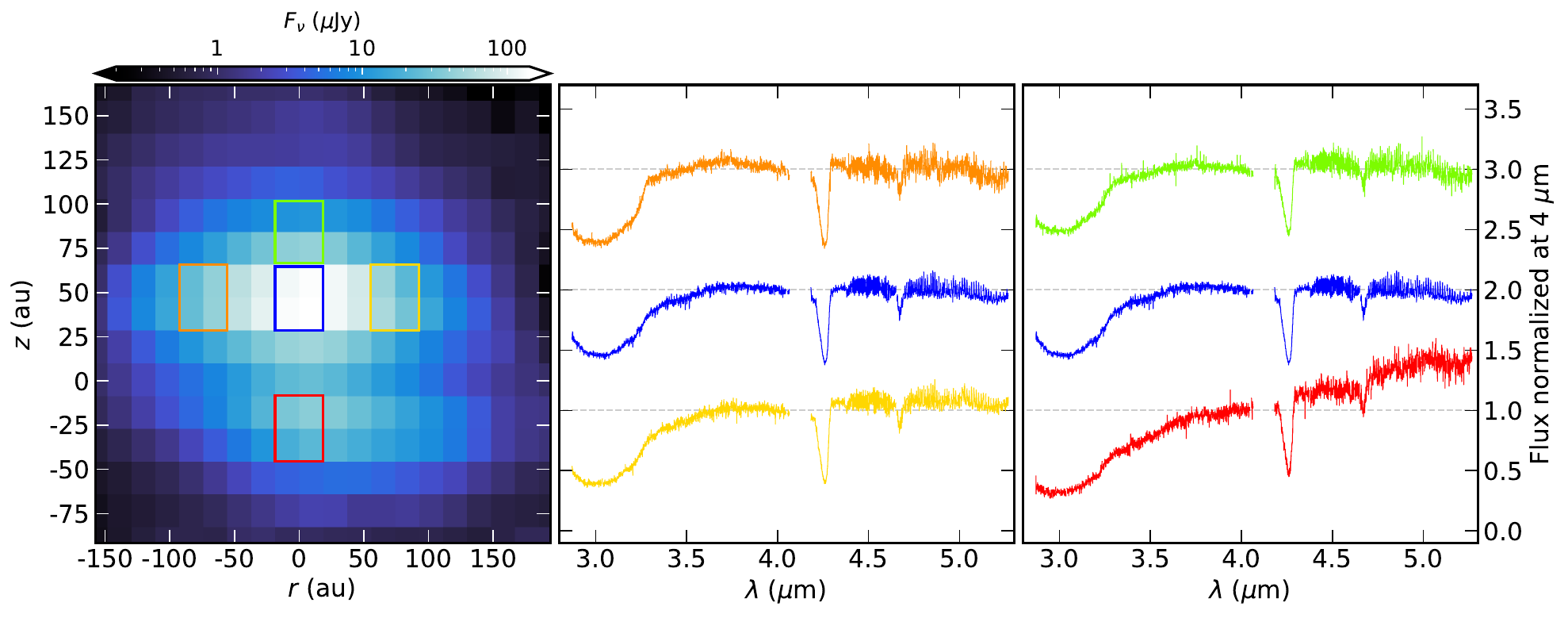}
    \captionsetup{format=hang}
    \caption{Spatial variations in the spectrum along the major and minor axes of the disk. \\
    \textbf{Left:} continuum flux of HH~48~NE at 4~\micron on a logarithmic scale. The colored squares mark the regions used to extract the spectra in corresponding colors in the middle and right panel.\\
    \textbf{Middle:} three spectra extracted along the major axis of the disk at radial distances of -70, 0 and 70~au with respect to the center. \\
    \textbf{Right:} three spectra extracted along the minor axis of the disk at heights of -25, 50 and 80~au with respect to the dark lane.     
    }
    \label{fig:regcomp spectra}
\end{figure*}
\begin{figure*}
    \centering
    \includegraphics[width = \linewidth]{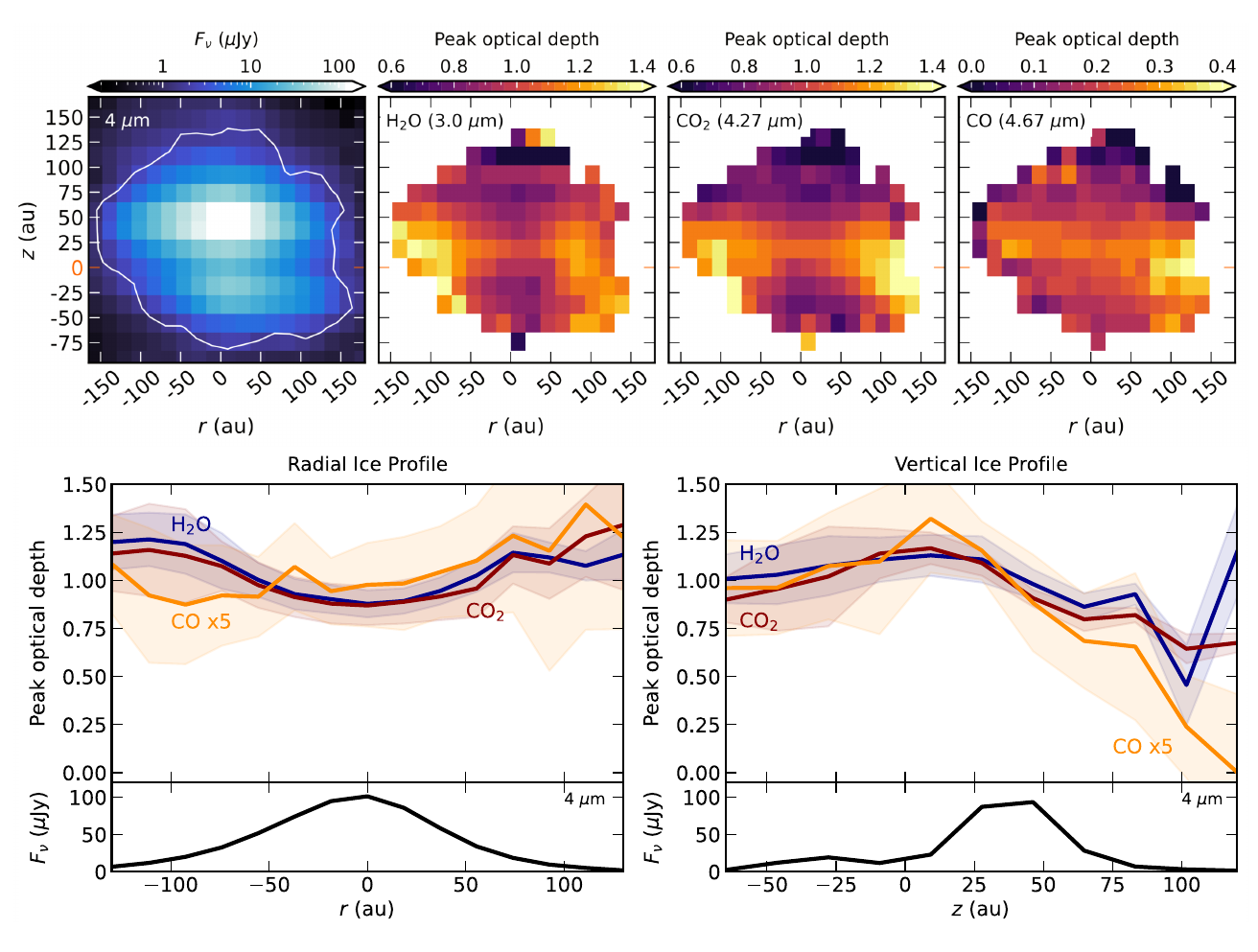}
    \captionsetup{format=hang}
    \caption{Overview of the variations in the relative strength of the three strongest ice features, \ce{H2O}, \ce{CO2}, and CO. \\
    \textbf{Top left panel:} continuum flux of HH~48~NE at 4~\micron on a logarithmic scale. The white line denotes the 5$\sigma$ detection limit on the continuum, determined in a featureless part of the spectrum, namely the 3.85~--~4~\micron region. This field of view is the same as the dashed rectangle in Fig.~\ref{fig:rgbim}\\
    \textbf{Other panels in the top row:} the peak optical depth per spaxel of \ce{H2O}, \ce{CO2}, and CO ice, respectively, determined using a Gaussian fit to the optical depth spectrum of each spaxel. \\
    \textbf{Bottom left panel:} median combined radial profile of the peak optical depth with the 1$\sigma$ spread around this line at different disk heights as a shaded region. The black line underneath shows the radial profile of the continuum at 4~\micron at $z$~=~50~au. \\
    \textbf{Bottom right panel:} median combined vertical profile of the peak optical depth with 1$\sigma$ spread around this line at different disk radii as a shaded region. The black line underneath shows the vertical profile of the continuum at 4~\micron at $r$~=~0~au.
    We note that $z$=0 is chosen as the dark lane between the upper and lower surface: in this frame, the star is located at $z\sim$10~au due to projection effects.}
    \label{fig:spatial dependence main ices}
\end{figure*}
\begin{figure*}[ht]
    \centering
    \includegraphics[trim = 2.2cm 0cm 1cm 0cm,clip,width = \linewidth]{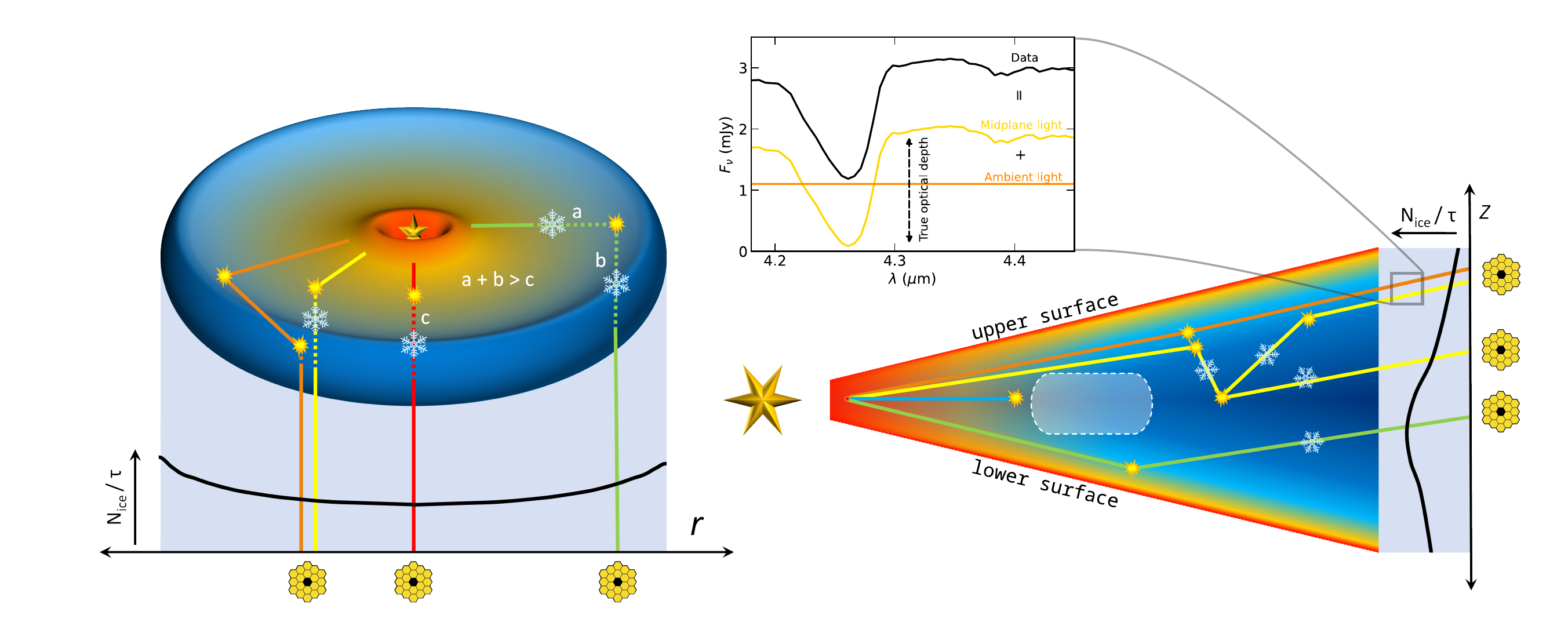}
    \captionsetup{format=hang}
    \caption{Cartoon of the physical origin of the radial (left) and vertical (right) profiles of the ice optical depths. \\
    \textbf{Left:} Since we observe ice absorption against the warm inner regions of a (close to) edge-on disk, all lines of sight trace ice along the full radial extent of the disk. This results in a largely constant optical depth profile along the radial direction. Since the path of the green photon trace is longer in the ice-rich outer disk (a + b) than the direct path (c), the traced ice column densities increase with radial separation of the observation.\\
    \textbf{Right:} Direct lines of sight (blue) towards the star and warm inner disk are blocked by the optically thick disk midplane (white region). The observed light follows therefore a specific path through the optically thin disk regions, including one or multiple scattering events. The flux in a JWST spaxel is composed of photons with different paths (orange and yellow), that have not necessarily crossed equal lengths in the cold parts of the disk. This means that photon paths that cross large parts of the ice-rich regions of the disk can be saturated locally, even though the optical depth of the combined contributions is a defined, small number (see inset). The optical depth increases towards the midplane, because the light paths through the cold disk layers are more likely compared to paths higher up in the disk (green). Since the scattered light from the lower surface is tilted towards the cold midplane, and similarly, light from the upper surface is tilted away from the cold midplane, the profile is asymmetric with deeper ice features in the lower surface.\\
    \textbf{Inset:} Qualitative schematic of the different flux contributions in the spectrum. The spectrum always has a contribution from light that passes through little ice (ambient light continuum; orange line) and light that passes through the cold regions of the disk, usually requiring multiple scattering events along the way (yellow line). The ratio of the contributions determines the optical depth in the observations if the feature is saturated. Photons with a single scattering event are less likely to reach regions with ice in the disk than photons with multiple scattering events \citep[see also Fig.~3 in][]{PaperII}.}
    \label{fig:cartoon}
\end{figure*}

\subsection{Spatial variation in ice features}\label{ssec: results - spatial distribution}
Since the disk is resolved in both vertical and the radial directions, we can map the strength of the ice features spatially for the first time in a protoplanetary disk.
In Fig.~\ref{fig:regcomp spectra} we present three normalized spectra across the disk along the major axis (middle panel) and the minor axis (right panel).
The overall structure of the spectrum is similar at all locations of the disk.
Along both the major and the minor axis, there is no big difference in the ice features and in the CO ro-vibrational lines. 
Spectra extracted further away from the central source along the major axis show a steeper slope in the water ice feature at $\sim$3.3~\micron, which may indicate that the relative amount of PAH emission increases in the outer disk.
The spectrum extracted from the lower surface shows an increased slope of the continuum, likely as a result of the radiative transfer effects for the disk that has an inclination of 83\degr.

We fit the same continua as for the disk-integrated spectrum (see Table~\ref{tab:continuum points}) in each spaxel with a continuum signal >5$\sigma$, estimated between 3.85 -- 4~\micron. 
We subsequently fit a Gaussian for each absorption feature, leaving amplitude, position, and FWHM as free parameters to determine the ``peak optical depth'' or ice feature amplitude in every spaxel.
The resulting maps for the strong ice features (\ce{H2O}, \ce{CO2}, and CO) are presented in Fig.~\ref{fig:spatial dependence main ices}.
We note that the parameters $(r,z)$ denote projected horizontal and vertical coordinates on the sky rather than intrinsic properties of the disk. 
Since the disk is not exactly edge-on, but rather inclined at $\sim$83\degr \ \citep{PaperI}, the upper surface of the disk ($z>0$) is different from the lower surface ($z<0$) in terms of absolute flux.

The peak optical depth maps consistently show a butterfly pattern, with the optical depth increasing towards the midplane in the vertical direction and outwards in the radial direction. 
The magnitude of the variations is only minor (<50\%) considering the multiple orders of magnitude difference in the continuum flux (blue panel in the top left of Fig.~\ref{fig:spatial dependence main ices}) and dust column density along a pencil beam\footnote{A pencil beam is defined as the direction along a straight line at the position of every pixel, neglecting the fact that photons originate from the source in the center.} (note that the color bar is truncated at a peak optical depth of 0.6 for \ce{H2O} and \ce{CO2}). 
There are some potential east-west asymmetries in the disk's lower surface, that could be mainly attributed to a higher continuum flux on the west side, which means that we can trace the ice features further out on that side (see discussion in beginning of Sect.~\ref{sec:Results}). 
Within the range where continuum emission is detected towards the east, there is no significant difference between the east and west side in any of the ice features.
Overall, vertical and radial cuts through the disk follow similar trends everywhere in the disk, but vary independent from each other.
In the subsequent sections, we will describe the variations in optical depth with the height and radius of the disk independently. 

\subsubsection{Radial variations}\label{sssec: results - spatial distribution - radial variations}
To highlight the radial trend in the ice optical depth, we took the median of the peak optical depth along the major axis of the disk, see bottom left panel of Fig.~\ref{fig:spatial dependence main ices}.
The 1~$\sigma$ variations on this radial profile across the vertical extent of the disk are marked as a shaded region around the median for reference.
Surprisingly, the shape is similar throughout the vertical extent of the disk.
The overall trend is that the optical depth increases with radius with an increase by 20\% at 100~au with respect to the center.
\ce{H2O}, \ce{CO2} and CO show identical spatial profiles within the uncertainty, with CO a constant factor of $\sim$5 weaker relative to \ce{H2O} and \ce{CO2}.
The ice absorption is in general symmetric in both sides of the disk.
The slight asymmetry in the CO profile could be a result of the low $S/N$ and the low number of pixels per bin at larger distances (see top right panel).
The radial profile cannot be directly translated into a radially resolved column density profile due to complex radiation transfer, including the fact that because of the edge-on geometry of the disk all sightlines sample the full radial extent of the disk. 
We will return to this point in Sect.~\ref{ssec:Discussion - ice in disk atmosphere}.
 
\subsubsection{Vertical variations}\label{sssec: results - spatial distribution - vertical variations}
To highlight the vertical variations in the ice optical depth, we took the median of the peak optical depth along the minor axis of the disk (see bottom right panel of Fig.~\ref{fig:spatial dependence main ices}). 
The 1~$\sigma$ variations on this profile across the radial extent of the disk are marked as a shaded region around the median.
We note that the shape is similar throughout the radial extent of the disk.
The overall trend in the three major ice species is that the vertical profiles peak around the dark lane and decrease quicker on the upper surface than on the lower surface.
We note that the origin is taken as the dark lane, which coincides with the highest peak optical depths, but the stellar position could likely be slightly above the dark lane due to projection effects.
The three major ice species (H$_2$O, CO$_2$ and CO) have identical profiles within the uncertainty from -70~au in the lower surface till 80~au in the upper surface.
We will return to the interpretation of this result in Sect.~\ref{ssec:Discussion - ice in disk atmosphere}.

\section{Discussion}\label{sec:Discussion}
Observations of ices in protoplanetary disks are crucial to understanding the chemistry that sets the composition of planetesimals and planetary atmospheres, and we show that with JWST we can trace the ice directly in late-stage edge-on disks.
Traditionally, the observed ices were believed to come from the regions that were simultaneously cold enough for particular ice species to freeze out but also sufficiently high up in the optically thin disk atmosphere to absorb stellar light scattered into the outer disk. 
Under these assumptions, the column densities of these ices would have provided relative measurements of CHONS-rich ice abundances in the disk layers just above the planet-forming midplane.
However, the observations presented in this work reveal a low \ce{^12CO2}/\ce{^13CO2} ratio, suggesting that the \ce{^12CO2} feature is saturated, even though the minimum flux is significantly above JWST/NIRSpec's sensitivity limit.
This raises the question of whether ratios between column densities, calculated from the observed optical depths using Eq.~\ref{eq:column density from band strength}, are at all meaningful proxies for relative ice abundances. 
Additionally, contrary to our expectations, only minor spatial variations are observed in the strength of the ice absorption features. 
This leads to a second question of which disk region(s) the observations are really sampling. 
In the subsequent sections, we explore which fundamental constraints on the answers to these questions can be extracted from such spatially resolved observations in a model-independent fashion.

\subsection{Where in the disk do these ice features originate?}\label{ssec:discussion - sampled region}
The radiative transfer in edge-on disks is complicated, as only very specific light paths reach the observatory.
In contrast to the pencil beam scenario, e.g. background star observations behind a dark cloud, all of the observed photons come from a similar region in the warm inner disk, before they are scattered into our line of sight \citep[see][for a deconstruction of the continuum]{PaperII}.
Fig.~\ref{fig:cartoon} illustrates examples of the photon propagation that could explain the observed absorption features and spatial variations. 
Most of the photons emitted by the star and the warm inner disk are absorbed along the way or scattered out of the line-of-sight.
Therefore, the observed ice absorption features may represent {\it both} the optically thin regions above and beyond the optically thick disk midplane (see right panel Fig.~\ref{fig:cartoon}) {\it and} potentially some part of the disk midplane itself, depending on the number of times each photon is scattered. 
Since the disk is axisymmetric, this means that all of the light that we observe has traveled through a similar radial extent of the disk, independent of the projected disk radius at which the light appears on the 2D JWST/NIRSpec array.
The slight increase of optical depth with radius (seen in the bottom left panel of Fig.~\ref{fig:spatial dependence main ices}) can be understood as a geometric effect: the further away we look from the star, the longer the light has traveled on average through the disk which effectively increases the ice column density along the light path (path a + b > c in the left panel of Fig.~\ref{fig:cartoon}) and the more likely the light has encountered ice along the way.

The light that we observe at every height in the disk took a complex path with one or more scattering events in-between (right panel of Fig.~\ref{fig:cartoon}).
The closer the light appears to come from the outer disk midplane (the dark lane at $z=0$~au), the greater the chance that the light has scattered through cold parts of the disk. 
This explains why we see a peak in the optical depth values at a height of $\sim$10~au above the midplane (bottom right panel of Fig.~\ref{fig:spatial dependence main ices}).
Since the disk is not exactly edge-on, but rather inclined at $\sim$83\degr, light that has traveled through the lower surface has a longer path through the cold midplane compared to light that traveled through the upper surface.
This results in slightly elevated optical depths on the lower surface compared to the upper surface, as seen in the bottom right panel of Fig.~\ref{fig:spatial dependence main ices}.
A final important result from Fig.~\ref{fig:spatial dependence main ices} is that there are no clear vertical cut-offs - or snowlines - visible in the ice feature depths, indicating that ices may extend up to 100~au in the disk, as discussed in Sect.~\ref{ssec:Discussion - ice in disk atmosphere}.

\subsection{Optical depths do not trace total ice column in outer disk}\label{ssec:Discussion - column densities}
Our data analysis suggests that ice feature optical depths, and by extension formal column densities, in protoplanetary disks, cannot be used to derive relative ice abundances in a straight-forward way.
We have illustrated in the previous section (Sect.~\ref{ssec:discussion - sampled region}) that the observed absorption features are likely not representative of the full extent of the disk, but trace only optically thin regions above the midplane and in the outer disk.
In addition, multiple light paths are combined in one line of sight, which implies that each spaxel contains multiple light contributions with different levels of ice absorption (see the inset panel of Fig.~\ref{fig:cartoon} for an illustration of this concept).
This means that there may be an ambient-light continuum underneath the ice features with much less ice absorption that fills in the ice absorption features, which has important consequences for the data analysis.
\ce{CO2}, \ce{H2O}, and potentially CO are locally saturated along the light path, indicating that light scattered through the cold parts in the disk is saturated in the ice absorption features (yellow line in the inset panel of Fig.~\ref{fig:cartoon}). 
The ice optical depth is then fully determined by the {\it fraction} of light scattered through the icy regions in the disk, relative to the light scattered through the ice-free regions (i.e., the ratio of orange to yellow flux in the inset panel of Fig.~\ref{fig:cartoon}); {\it the measured optical depths do not represent the total column density of ice along the light path.}

Increasing the amount of ice in the cold regions of the disk will not have a significant effect on the measured optical depth: the depth of the ice feature in the inset of Fig.~\ref{fig:cartoon} is determined primarily by the level of the orange line rather than the depth of the yellow line.
This effect was already predicted using radiative transfer models of HH~48~NE in \citet{PaperII}, where we studied the sensitivity of the ice features on the physical disk parameters.
We used a model tailored to HH~48~NE with constant ice abundances throughout the disk (taken as 80, 22, and 99~ppm for \ce{H2O}, \ce{CO2}, and CO respectively), assuming that all molecules are frozen out below their desorption temperature and below a UV threshold of A$_\mathrm{v}$ = 1.5. 
Varying the abundances in the model, we find that similar peak optical depth values of $\sim$0.9-1.2 for \ce{H2O} and \ce{CO2} in HH~48~NE models require abundances of 2~--~100 times the assumed ISM values: high enough to saturate these ice absorption features locally in the disk.
An accurate column density can only be determined if we know what fraction of the light is scattered through regions without ice in the disk; i.e. we must determine the ambient light flux indicated by the orange line in the inset of Fig.~\ref{fig:cartoon}, which yields the zero point for the flux of the yellow line. 
The radiative transfer models of HH~48~NE show that photons with a single scattering event - and therefore less ice absorption than photons with multiple scattering events - account for $\sim$1/3 of the flux at NIRSpec wavelengths, which is comparable with the ambient light contribution observed \citep[see Fig.~3 in][]{PaperII}.
The fraction of ambient scattered light is impossible to know without modeling the radiative transfer through the disk.
Therefore column densities derived using Eq.~\ref{eq:column density from band strength} should be interpreted as lower limits.

An important implication of this is that naive ice column density ratios, calculated by multiplying the observed integrated optical depth in source integrated spectra with the band strength, do not necessarily correspond to similar relative ice abundance ratios in the disk.
The region where molecules are frozen out is different for every molecule, and therefore there are also differences between the ratio of ambient scattered light to scattered light that encounters locally high column densities of each ice species.
Since the ambient light continuum is similar for the \ce{CO2} and \ce{H2O} bands - due to their proximity in wavelength, and their similar snowline location in the disk - this results in comparable optical depths in the two features ($\sim$0.9), even though their actual relative abundance may be very different.
The ratio of light that reaches the ices near the midplane will be even more different for volatile species like CO.

With these caveats in mind, we calculate formal column densities, presenting them as lower limits, to enable comparison with literature observations.
The lower limits on the column of \ce{H2O}, \ce{CO2}, and CO are >1.9~$\times~10^{18}~\mathrm{cm}^{-2}$, $>$2.1~$\times10^{17}~\mathrm{cm}^{-2}$ and $>$2.3~$\times~10^{17}~\mathrm{cm}^{-2}$, respectively.
The detection of \ce{^13CO2} with a column density of $>$1.3~$\times~10^{16}~\mathrm{cm}^{-2}$ is crucial, as this allows us to determine the actual \ce{CO2} column density, using ISM assumption for the $^{12}$C/$^{13}$C isotope ratio.
Since the main isotopologue is locally saturated and we know that \ce{^12CO2} and \ce{^13CO2} should be located in the same region in the disk, we can correct for the ice-free ambient-light continuum for the \ce{^13CO2} feature by removing a constant flux (orange line in the inset panel of Fig.~\ref{fig:cartoon}) underneath the peak of the \ce{CO2} feature.
Removing the ice-free ambient-light continuum results in an increase of the peak optical depth of \ce{^13CO2} from 0.06 to 0.010 and an increase in the integrated optical depth by a factor of 1.6.
The estimated \ce{^13CO2} column density using this approach is 2.1$~\times~10^{16}~\mathrm{cm}^{-2}$.
If we assume the ISM \ce{^12C/^13C} ratio of \citep[77][]{Wilson1999_13Cratio}, then we can find an estimate of the actual \ce{^12CO2} column density of 1.6$~\times~$10$^{18}$~cm$^{-2}$.

We note that this estimated column density is one order of magnitude higher than the lower limit found from multiplication of the observed integrated optical depth of the \ce{^12CO2} feature with the band strength, illustrating the importance of detecting the less abundant isotopologues with JWST.
Inferring molecular abundances from the main isotopologue ice features, as traditionally done in the literature, may drastically underestimate the amount of ice present and should be avoided.
Therefore, it is critical to observe either less optically thick, rare isotopologues like HDO, \ce{^13CO2} and \ce{^13CO} or a disk of sufficiently low mass in order to determine the correct order of magnitude for the column densities of ices in the comet-forming regions of disks.

\subsection{Excess ice in disk atmosphere?}\label{ssec:Discussion - ice in disk atmosphere}
\begin{figure}[!b]
    \centering
    \includegraphics[width = \linewidth]{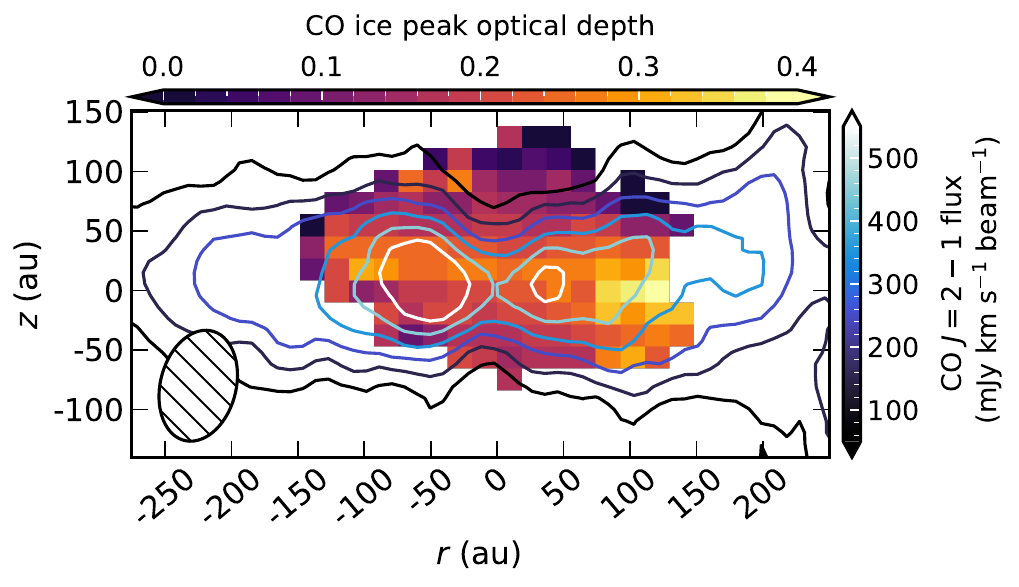}
    \caption{Spatial variations in the CO ice peak optical depth (colorscale) compared to ALMA CO~$J=2-1$ observations presented in \citet{PaperI} (contours).
    The beam of the ALMA observations (0\farcs{51}~$\times$~0\farcs{31}) is shown in the bottom left corner. 
    The JWST/NIRSpec observations have an effective resolution of 28~au or 1.5~spaxels.}
    \label{fig:CO ice compared to ALMA gas}
\end{figure}

Assuming that the ice absorption takes place in the optically thin disk outer regions, the vertical snow surfaces caused by thermal desorption and photo-desorption should produce a noticeable drop in the peak optical depth with disk height ($z$ in Fig.~\ref{fig:spatial dependence main ices}), as explained in Sect.~\ref{ssec:discussion - sampled region}.
However, the three most prominent ice features \ce{H2O}, \ce{CO2}, and CO extend vertically over 125~au with only a minor decrease in strength.
This is especially curious for CO ice, which is thought to sublimate thermally at $\sim$20~K in protoplanetary disks \citep[][and references therein]{Collings2003,Pinte2018,Minissale2022} and is therefore limited to about one hydrostatic scale height, which lies at $\sim$$z/r$~0.15 (\citealt{Pinte2018}, \citealt{Law21_MAPSIV} and see \citealt{PaperII}). 
The observed CO ice is even seen at similar maximum height as the CO~$J=2-1$ gas observations reported in \citet{PaperI} (see Fig.~\ref{fig:CO ice compared to ALMA gas}), which corresponds to temperatures of >50~K.

There are four potential scenarios that could result in a similar resolved profile for the three main ice features. 
Here, we describe the possibilities and identify the one that is the most viable, given what is known about this particular source.

First, the extra absorption provided by a foreground cloud or envelope with high ice column densities could explain the relatively constant peak optical depths. 
In that case, the whole disk would function as the light source, and all the emission that is scattered from the disk would pass through a similar cloud, which would give a relatively similar ice absorption at each spaxel across the disk. 
However, this option is unlikely given the measured visual extinction towards stars in the neighbourhood ($A_\mathrm{V}$~$\sim$3) by GAIA \citep{GAIADR3} and the low visual extinction derived from the constraints on the disk geometry and stellar properties \citep[$A_\mathrm{V}$~$\sim$5][]{PaperI}. 
The integrated optical depths of the ice features (Table~\ref{tab:continuum points}) are a factor of 5 above the observed trend with $A_\mathrm{V}$ towards background stars \citep{Boogert2015}, which means that a potential foreground cloud contributes <20\% of the total depth to the ice absorption. 
Additionally, we find that the absorption features of \ce{CO2} and \ce{H2O} are saturated at a flux >0, which is only possible if absorption in the disk dominates. 
If a foreground cloud dominates the absorption, the absorption features would saturate at 0 due to the lack of ice-free ambient-light continuum.
The observed optical depths at $z~=~125$~au are therefore likely not due to a foreground cloud.

Second, the observed profile could be a result of radiative transfer in the disk.
If most of the ice absorption happens in regions relatively close to the star, and the observed features are scattered through the disk in a similar way as the continuum, the vertical distribution could be uniform. 
This is, however, not in line with our current knowledge of the temperature structure of the disk \citep{PaperI}, because the CO snowline is estimated to be at $r \sim$50~au, comparable to observed snowlines in other systems \citep{Qi2013_CO_snowline,Vanthoff2020,Podio2020,Zhang2021_maps_COstructure} and the scattering of the ice band should therefore be inherently different than the continuum light from the warm inner disk and the \ce{CO2} and \ce{H2O} features. 
Alternatively, light could have previously been scattered down below the CO snow surface, before scattering into the disk atmosphere and then again into the line-of-sight towards us (yellow line in the right panel of Fig.~\ref{fig:cartoon}). 
However, radiative transfer models including such anisotropic scattering show layered disk ice segregation as a result of temperature gradients and photo-desorption \citep{Ballering2021,Arabhavi2022,PaperII}.
Even though photon packages may have encountered multiple scattering events along the light path \citep[see also][]{PaperII}, there is no reason to assume that scattering to deeper layers is more efficient than scattering through the upper layers alone, which would still result in noticeable vertical differences. 

Third, a fraction of the CO ice could be trapped in ice with a much higher desorption temperature, and therefore survive at much higher temperatures than that expected of a pure CO ice. 
Laboratory experiments of CO and \ce{CO2} desorption with \ce{H2O} show that CO deposited on top of a \ce{H2O} ice base layer can diffuse into the \ce{H2O} matrix, remaining trapped due to structural changes of the \ce{H2O} ice while heated, and effectively co-desorbs at the crystallization temperature of \ce{H2O} \citep[150~K; e.g.,][]{Collings2003,Palumbo2006,Fayolle2011,Lauck2015}.
Similar diffusion is also seen for CO deposited onto \ce{CO2} \citep{Palumbo2006,Simon2019}, where the diffusion into \ce{CO2} is more efficient than into \ce{H2O}, and CO can also be trapped in a \ce{CH3OH} ice matrix \citep{Ligterink2018}. 
Entrapment of part of the CO in ices with a higher desorption temperature would still be in line with observations of gas-phase CO that show that most of the CO desorbs at $\sim$20~K, given that the features are saturated (see the discussion in Sect.~\ref{ssec:Discussion - column densities}).
The hypothesis that CO is trapped in the ice matrix is supported by the width of the CO feature, which is broader than that of pure CO, and also by shifts in the peak wavelengths of the \ce{^12CO2} and \ce{^13CO2} features, which can occur due to a \ce{H2O} or \ce{CO} rich environment. 
Such profile signatures were seen by \citet{McClure2023} in the CO and \ce{CO2} ice towards background stars in the same star-forming region, and ultimately found to originate in mixed ices, after comparison with laboratory data. 
Due to the complicated and extended disk geometry, a detailed retrieval of the ice mixtures from the profiles must be done in conjunction with a full radiative transfer model, which we reserve for a future paper.

The trapping of CO in mixed ices would not explain the lack of snow surfaces for \ce{CO2} and \ce{H2O}, though, unless there is a mechanism for those ices to remain trapped in a less volatile material on the grain surface. 
However, chemical modeling including vertical turbulent mixing and diffusion shows that icy grains can reach the observable disk surface in turbulent disks, effectively mixing ices upward faster than they can be photo-desorbed in the upper layers \citep{Semenov2006,Semenov2008,Furuya2013,Furuya2014,Woitke2022}.
The grain sizes of the icy grains inferred from the \ce{H2O} and \ce{CO2} features (>1\micron; see Sect. \ref{ssec: Observations - ice features}), that are thought to trace mainly the disk atmosphere, could be a direct result of mixing ice rich micron-sized dust grains to the disk upper layers.
While the timescales for thermal desorption are short, this could potentially explain the lack of a clear \ce{CO2} and \ce{H2O} snowsurface. 
Regardless of the mechanism, if it is confirmed that a fraction of the ice survives up to the disk surface, this would have major consequences on the disk chemistry and models of planet formation, as we discuss in Section~\ref{ssec:Discussion - chemical implications} below. 

\subsection{Implications for the outer disk chemistry}\label{ssec:Discussion - chemical implications}
Using the initial ice radiative transfer models of HH~48~NE presented in \citet{PaperII} and described in more detail in Sect.~\ref{ssec:Discussion - column densities}, a first order estimate of the molecular abundance can be derived using the relation between the peak optical depth of the absorption feature and the molecular abundance. 
We find that abundances should be significantly higher than the ISM abundance (taken as 80, 22, and 99~ppm for \ce{H2O}, \ce{CO2}, and CO respectively) in order to reproduce the observed optical depths of the three main ice species and the level of saturation traced with the \ce{^13CO2} ice observation.
Additionally, the observed ice features are stronger than anticipated based on current chemical models \citep[e.g.][]{Ballering2021,Arabhavi2022}. 
This could mean that ices are efficiently inherited from the pre-stellar cloud stage \citep{Ballering2021}.
However, the results suggest that there is even more ice (>2$\times$) than can be explained by inheritance from the pre-stellar phase alone.

If more ices are frozen out at the disk midplane than expected, this could impact the interpretation of the CO, C, and \ce{H2O} gas depletion seen in the outer regions of some disks with ALMA \citep[e.g.,][]{Van_Zadelhoff2001, Miotello2017, Sturm2023lkca15, Du2017_h2o_abu, Van_dishoeck2021}.
Efficient conversion from volatile gas to the ice phase has long been suggested as an explanation for the CO and \ce{H2O} depletion in the outer disk gas, requiring a combination of freezeout, chemical conversion, grain growth, settling, radial drift, and vertical mixing \citep{Bosman2018,Schwarz2018,Krijt2018,Krijt2020}.
The non-detection of \ce{CH3OH} in our data suggests that, if chemical conversion is the primary depletion mechanism, CO is more likely converted into \ce{CO2} than in more complex molecules like \ce{CH3OH}.

The co-existence of CO ice in the emitting layer of CO gas due to entrapment in less volatile ices, as discussed in Sect.~\ref{ssec:Discussion - ice in disk atmosphere}, could be a another mechanism contributing to the low CO abundances measured in the gas phase of a large population of T Tauri stars. 
However, lab measurements suggest that entrapment could only result in CO gas depletion by less than a factor of 2 \citep{Simon2019}, which is insufficient to explain the 2 orders of magnitude depletion observed in older systems like TW Hya and DL Tau \citep{Kama2016TWHya,Sturm2022CI}; in several T Tauri systems, the inner disk is also depleted in atomic C, which is still best explained by efficient trapping of icy dust grains at the edges of disk gaps \citep{mcclure2019,mcclure2020}. 

Determining which of these mechanisms is operating in HH~48~NE will be essential for procedures that measure disk gas masses using volatiles like CO, as the CO gas mass is frequently used as one proxy for the bulk H$_2$ gas mass in disk surveys, in the absence of better tracers like HD \citep{bergin2013,mcclure2016,Sturm2023lkca15}.
As one of the most fundamental properties of disks around young stars, good constraints on the total gas mass are crucial to understanding disk evolution and the potential of any given disk to form planets.

Finally, the observed low \ce{^12CO2}/\ce{^13CO2} ratio of 14 compared to the ISM ratio of 77 \citep{Wilson1999_13Cratio}, is interpreted in Sect.~\ref{ssec:Discussion - column densities} to be likely a result of local saturation of the main feature. 
However, isotopic ratios in protoplanetary disks can be used as tracers of chemical fractionation processes during planet formation that are reflected in observed planetary atmospheres \citep[see e.g.,][]{Zhang2021_13coratio_hotjupiter,Yoshida2022}.
If future radiative transfer modeling suggests that the \ce{^12CO2}/\ce{^13CO2} isotopic ratio is genuinely lower than in the ISM, rather than  deriving from local saturation effects, then this could indicate fractionation or isotope-selective destruction processes at work in this disk \citep{Boogert2000}. 
In that event, edge-on disks could represent an exciting new avenue to explore solid-state fractionation in a way that is comparable to sample return missions to Solar System bodies.

\subsection{Comparison with previous observations}\label{ssec:Discussion - previous observations}
In this paper, we present the first detection of \ce{^13CO2} ice in a Class~II disk, along with the first firm localization of CO ice to the disk, and the tentative identification of trace ice species \ce{OCN^-}, \ce{OCS}, and \ce{NH3}.  
While CO ice was previously detected towards a disk line of sight with Very Large Telescope (VLT) \citep{Thi2002}, it was later shown that the majority of the CO ice must be present in intervening cloud material \citep{Pontoppidan2005}.  
With AKARI, CO and \ce{^13CO} ice were only seen towards younger, embedded sources \citep{Aikawa2012}.

A qualitative comparison with previous observations reveals that the near-IR spectrum of HH~48~NE is remarkably similar to IRAS~04302, an edge-on Class I protostar, but bears less resemblance to the other sources observed by AKARI \citep{Aikawa2012}.  
This includes the absorption signatures near 2.9 and 3.1~\micron, seen in HH~48~NE and IRAS~04302, but not in other systems, and the extent of the scattering versus hydrate contribution in the 3~\micron red wing.
The red scattering wing and relative optical depths of the \ce{CO2} and CO features are also similar. 
This similarity is somewhat surprising given that IRAS~04302 has an additional envelope component \citep{Wolf2003} that is not seen in HH~48~NE \citep{PaperI}. 
It may suggest that the envelope of IRAS~04302 is tenuous enough not to contribute to the near-infrared ice features, or it may indicate that the envelope ices in Class~I sources are similar to disk ices in Class~II sources. 

\subsection{Future prospects and outlook}\label{ssec:Discussion - future prospects}
Clearly, observations of additional edge-on systems are needed to understand the relative importance of intrinsic disk properties vs. ~viewing geometry in setting the near-IR spectrum. 
Including more sources in the sample will help to constrain which properties are inherent to protoplanetary disks, and which should be attributed to source-specific properties like age, envelope material, and dynamical interaction with a binary component.
Future comparison of HH~48~NE's near-infrared with the comparable NIRSpec spectrum from the neighboring edge-on protostar Ced~110~IRS~4, also observed as part of the Ice Age JWST ERS program, will enable us to quantify the similarity of ices between the Class~I and Class~II stages.

Additionally, many ice species discussed in this work have corresponding absorption features in the JWST/MIRI wavelength range, 5~--~28~\micron. 
Longer wavelengths may trace different regions in the disk due to differences in scattering and there could be a significant contribution from direct thermal dust emission at longer wavelengths \citep[see Fig.~A.1. in][]{PaperII}.
However, a future comparison between NIRSpec and MIRI observations will help to constrain the strength of the potential PAH emission feature at 3.28~\micron, since that feature correlates with the features at 6 and 8~\micron.
Additionally, it could confirm our conclusions about the presence of \ce{NH3} and lack of \ce{CH3OH}, as both of these species have stronger absorption features at 9.0 and 9.7~\micron, respectively.

In this work, we have clearly demonstrated that neglecting the role of radiative transfer in the interpretation of ice observations towards sources with substantial physical structure, like edge-on protoplanetary disks, can lead to incorrect abundances that could be off by an order of magnitude, in the case of \ce{CO2}. 
Due to the radiative transfer, absorption feature shapes and depths are affected, which complicates determination of abundance ratios and ice environments.
In forthcoming papers, we will study the degree to which the effect of the radiative transfer on the feature shape can be disentangled from the effects of the chemical environment of the ices, based on the results of this paper.
Additionally, we will use these observations to improve our already existing ice model \citep{PaperII} to study the spatial variations in ice optical depth and molecular abundances in more detail and derive an ice to silicate ratio using the additional JWST/MIRI spectra.

\section{Conclusions}\label{sec:Conclusion}
We presented the first spatially resolved observations of the main ice species in a mature, envelope-free edge-on Class II protoplanetary disk, HH~48~NE, using JWST/NIRSpec. 
These results demonstrate the novel capabilities of JWST for revealing the ice inventories of disks. 
After analyzing the observed ice features and spatially resolved trends, we can conclude the following:
\begin{itemize}
    \item A variety of ice features are detected including the major ice components \ce{H2O}, \ce{CO2}, and CO and multiple weaker signatures from less abundant ices,  \ce{NH3}, \ce{OCN^-}, and OCS. Additionally, we detect the isotopologue \ce{^13CO2} ice absorption feature for the first time in a protoplanetary disk.
    \item The combination of an integrated absorption ratio of \ce{^12CO2}/\ce{^13CO2}=14, in contrast to the ISM value of 77, plus similar peak optical depths of \ce{H2O} and \ce{CO2} indicates that the main \ce{^12CO2} and \ce{H2O} features are saturated, despite excellent S/N at the bottom of both features. Their saturation implies that the spectrum has a strong contribution from an ice-free ambient scattered-light continuum, which reduces the apparent optical depth, despite the high ice column densities encountered by light paths through the cold disk regions. Using the \ce{^13CO2} absorption feature, we show that the naive calculation of column densities from the observed integrated optical depth of the \ce{^12CO2} feature results in an underestimation of the total ice column in the disk by more than one order of magnitude. {\it Detecting the less abundant isotopologues with JWST is therefore crucial for measuring molecular abundance ratios in the ice.}
    \item The shape of the crystallized water ice feature and the red shoulder of the \ce{CO2} feature indicate that a significant fraction of the probed dust grains along the photon path are larger than $\sim$1~\micron.
    \item Radial variations in ice composition and abundance are hard to trace in edge-on disks due to the complex radiative transfer effects, assuming that all light originates from the warm inner disk and travels through the entire radial extent at all observed projected radii on the sky. A slight increase in feature depth with projected radial separation from the star is interpreted as a result of an increased light path length through the cold regions in the outer disk. 
    \item Vertical variations in ice optical depth are also small and \ce{H2O}, \ce{CO2}, and CO reveal similar distributions, although these molecules have significantly different desorption temperatures. While a contribution from ices in a foreground cloud and complicated radiative transfer scattering making the ice features appear coming from higher layers cannot be excluded entirely, we find that CO ice likely survives higher up in the disk than expected based on their desorption temperature of $\sim$20~K. This may be due to the entrapment of CO in less volatile ices like \ce{H2O} and \ce{CO2}.
    \item The ice features are deeper than anticipated from radiative transfer modeling assuming typical interstellar ice abundances with respect to hydrogen.
    The high column densities of ice observed in HH~48~NE suggest that we see a result of gaseous CO and \ce{H2O} conversion to ices, confirming the impact of earlier proposed mechanisms to explain the observed low CO gas fluxes. If it is verified, by modeling of the ice profiles, that part of the CO ice is trapped in \ce{CO2} and \ce{H2O} ice and survives high up in the disk, then CO diffusion into the ice could also contribute to the CO depletion seen in the gas phase. 
\end{itemize}
Understanding the gas/ice ratio of important volatiles is crucial in our understanding of the chemistry in planet-forming regions and the physical properties of protoplanetary disks like the gas mass.
Many of the features of these resolved ice observations and their implications on the chemistry and physics of protoplanetary disks can be readily understood in a model-independent way, as demonstrated in this work. 
Future comparison over the full wavelength coverage of JWST, including MIRI data, will further constrain some of the ice abundances, as well as the ice/rock ratio. 
However, the spatial origin of some observed signatures remains uncertain and will require advanced radiative transfer modeling, as well as a larger sample of edge-on disk observations, to fully extract relevant information about the birthplace of planets and comets.

\begin{acknowledgements}
We thank the referee for comments that helped to improve the manuscript.
Astrochemistry in Leiden is supported by the Netherlands Research School for Astronomy (NOVA), by funding from the European Research Council (ERC) under the European Union’s Horizon 2020 research and innovation programme (grant agreement No. 101019751 MOLDISK).
M.K.M. acknowledges financial support from the Dutch Research Council (NWO; grant VI.Veni.192.241).
D.H. is supported by Center for Informatics and Computation in Astronomy (CICA) grant and grant number 110J0353I9 from the Ministry of Education of Taiwan. 
D.H. acknowledges support from the National Technology and Science Council of Taiwan through grant number 111B3005191.
E.D. and J.A.N. acknowledge support from French Programme National ‘Physique et Chimie du Milieu Interstellaire’ (PCMI) of the CNRS/INSU with the INC/INP, co-funded by the CEA and the CNES. 
M.N.D. acknowledges the Swiss National Science Foundation (SNSF) Ambizione grant no. 180079, the Center for Space and Habitability (CSH) Fellowship, and the IAU Gruber Foundation Fellowship.
Part of this research was carried out at the Jet Propulsion Laboratory, California Institute of Technology, under a contract with the National Aeronautics and Space Administration (80NM0018D0004).
S.I., E.F.vD, and H.L. acknowledge support from the Danish National Research Foundation through the Center of Excellence “InterCat” (Grant agreement no.: DNRF150).
M.A.C. was funded by NASA’s Fundamental Laboratory Research work package and NSF grant AST-2009253.
\end{acknowledgements}

\bibliographystyle{aa}
\bibliography{refs.bib}

\begin{appendix}
\section{CO ro-vibrational gas line fitting}
\begin{figure*}[!b]
    \includegraphics[width = .85\textwidth]{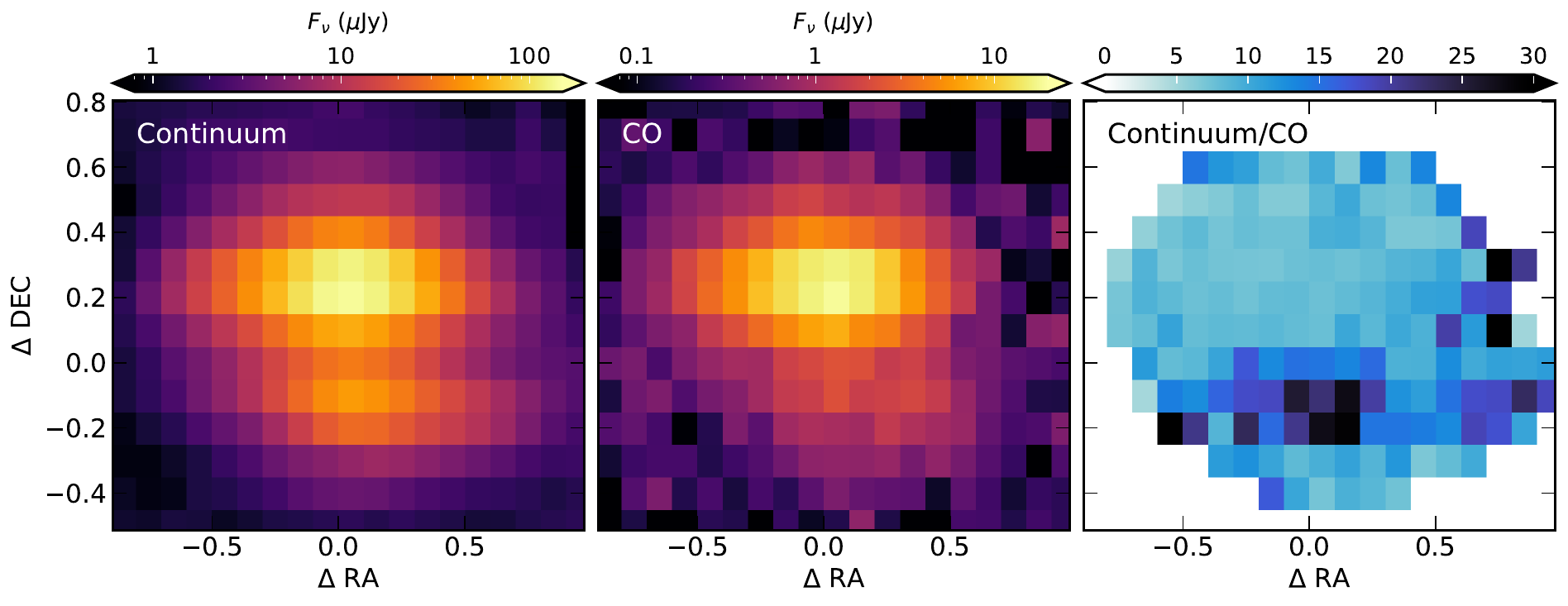}
    \vspace{-3mm}
    \centering
    \captionsetup{format=hang}
    \caption{Comparison between the spatial distribution of the CO emission and the continuum emission. \\
    \textbf{Left:} Spatial distribution of the median continuum flux in the region with CO gas emission (4.36~--~5.2~\micron).\\
    \textbf{Middle:} Spatial distribution of the median of the peak fluxes of the CO gas emission lines.\\
    \textbf{Right:} Ratio of the left two panels showing a constant ratio in the upper surface, but a higher ratio in the lower surface due to missing CO flux. }
    \label{fig: CO spat appendix}
\end{figure*}

\begin{figure}
    \centering
    \includegraphics[width = 1\linewidth]{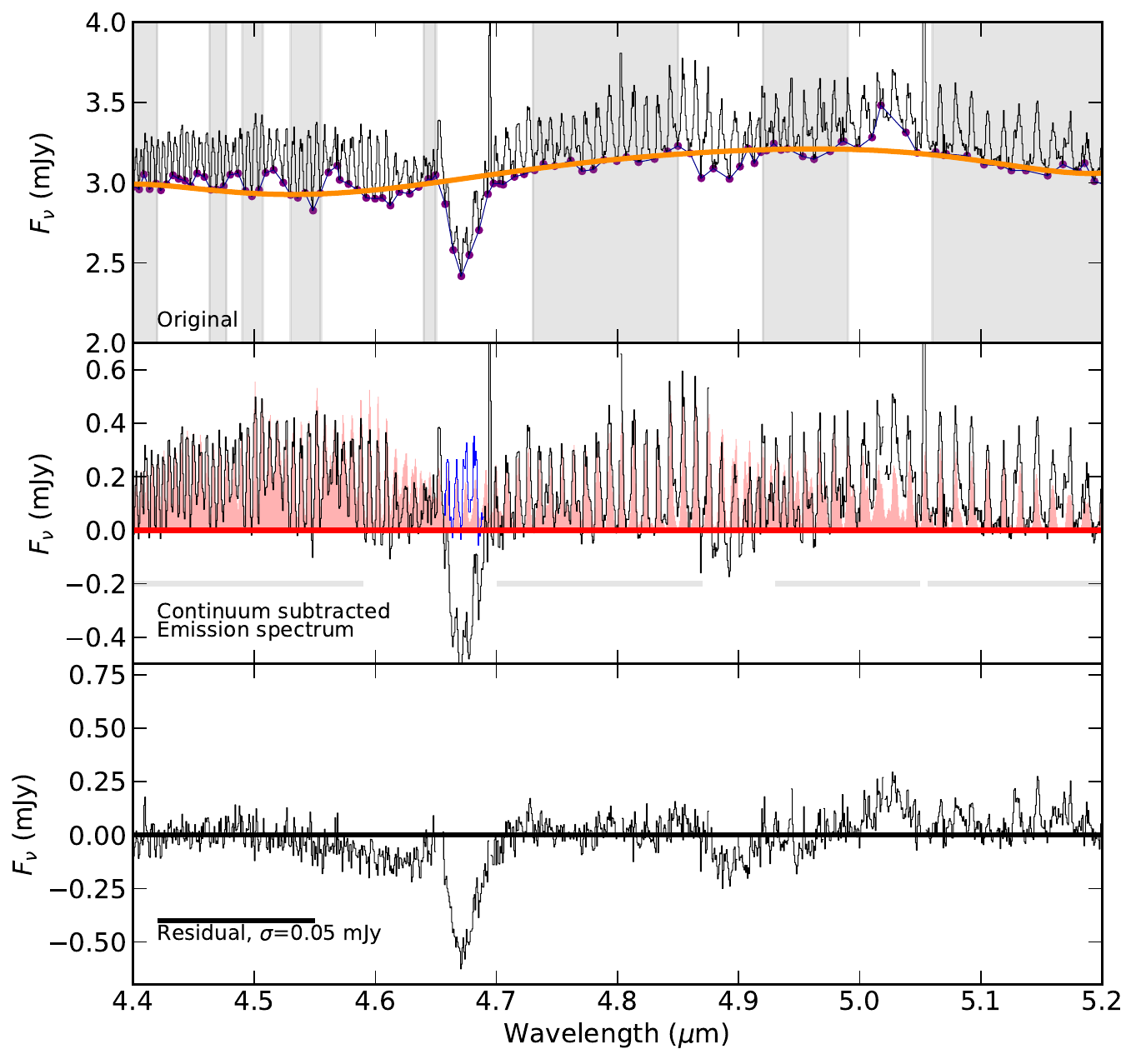}
    \captionsetup{format=hang}
    \caption{Overview of the CO gas line fitting procedure following the steps described in the text.\\
    \textbf{Top panel:} Observed source integrated spectrum of HH~48~NE with selected troughs in between the emission lines in purple. Regions used to fit the continuum underneath the lines are marked in gray.\\
    \textbf{Middle panel:} Continuum subtracted CO emission spectrum, corrected for the optical depth in the CO ice feature (blue region). The best fitting model is shown in red, regions taken into account in the fitting are marked with a gray horizontal line.\\
    \textbf{Bottom panel:} Residual spectrum of subtracting the model from the data. The standard deviation over this region is 0.05~mJy, compared to 0.27~mJy in the CO line-free continuum at $\sim$4~\micron (see Fig. \ref{fig:overview_figure}).}
    \label{fig: CO fit overview appendix}
\end{figure}

\label{app:co fitting}
The HH~48~NE observations include a forest of ro-vibrational CO lines from 4.3~\micron to 5.28~\micron that are spectrally overlapping and therefore create a quasi-continuum underneath the lines.  
This results in uncertainty on the continuum for the analysis of the ice absorption bands.
In an attempt to remove the gas lines, we followed a similar approach as in \citet{Grant2023} by fitting a LTE slab model to the observed spectrum with a constant temperature, column density and emitting area. 
First, we estimated the continuum by selecting regions in the brightest CO lines and selecting regions with the least amount of line overlap and ice absorption features (see the purple dots and grey areas in Fig.~\ref{fig: CO fit overview appendix}, respectively).
Next, we fitted LTE slab models to the lines in the regions marked with a grey horizontal line in the middle panel of Fig.~\ref{fig: CO fit overview appendix} using a grid of models varying the column density, excitation temperature and linewidth.
Since we observe an unknown small fraction of the CO line flux scattered through the outer disk, we normalize the fitted spectrum to the observed spectrum minimizing the reduced $\chi^2$ value.

The best-fitting model has an excitation temperature of 2070~K, a column density of 1.4$~\times~ 10^{19}$~cm$^{-2}$, with a line FWHM of 150~km~s$^{-1}$.
We note that these values result in a model that fits the CO emission lines well and improves the quality of the continuum, but that these values should not be trusted to imply physical quantities or regions of origin.
The radiative transfer can have a significant impact on the line ratios over the spectral region that is used for the line fitting.
We can conclude that the emitting area has to be warm (>1500~K) to fit the lines with high $E_{\rm up}$ and that the lines are spectrally resolved indicating high velocities.

The continuum-subtracted CO emission follows a similar spatial distribution as the continuum at similar wavelengths (see Fig.~\ref{fig: CO spat appendix}), which indicates that the CO likely comes from the obscured inner disk or jet launching region and is scattered through the outer disk in our line of sight in a similar way as the continuum.
The CO gas emission is less prominent in the lower surface, which indicates that the emitting region may be higher up than the emitting region of the continuum.
We used the ratio of the CO lines to the continuum (right panel of Fig.~\ref{fig: CO spat appendix}) to scale the best-fitting CO spectrum and remove the lines from the continuum in every individual spaxel.

\end{appendix}
\end{document}